\documentclass[twocolumn,showpacs,nofootinbib,aps,prd]{revtex4}
\usepackage{graphicx,dcolumn,bm,amsfonts,amsmath,color,xcolor,amsthm}
\usepackage[colorlinks=true, citecolor=blue, linkcolor=blue,urlcolor=blue]{hyperref}
\usepackage{slashed}
\usepackage{ulem}

\allowdisplaybreaks

\begin{document}
\title{Exploring the exclusive decay $B^+\to \omega\ell^+\nu$ with light-cone sum rules}
\author{Yin-Long Yang$^*$}
\author{Ya-Lin Song\footnote{These authors contributed equally to this work.}}
\author{Fang-Ping Peng}
\address{Department of Physics, Guizhou Minzu University, Guiyang 550025, P.R.China}
\author{Hai-Bing Fu}
\email{fuhb@gzmu.edu.cn}
\author{Tao Zhong}
\address{Department of Physics, Guizhou Minzu University, Guiyang 550025, P.R.China}
\address{Institute of High Energy Physics, Chinese Academy of Sciences, Beijing 100049, P.R.China}
\author{Samee Ullah}
\address{Institute of High Energy Physics, Chinese Academy of Sciences, Beijing 100049, P.R.China}
\address{University of Chinese Academy of Sciences, Beijing 100049, P.R.China}

\begin{abstract}
In this paper, we calculate the Cabibbo-Kobayashi-Maskawa matrix element $|V_{ub}|$ by the semileptonic decay $B^+\to \omega\ell^+\nu$. For the transition form factors (TFFs) $A_1(q^2)$, $A_2(q^2)$, and $V(q^2)$ of $B^+\to \omega$, we employ the QCD light-cone sum rules method for calculation, and by constructing the correlation function using left-handed chiral current, we make the $\delta^1$-order twist-2 light-cone distribution amplitudes (LCDA) $\phi^\| _{2;\omega}(x,\mu)$ dominate the contribution, in which the twist-2 LCDA $\phi^\| _{2;\omega}(x,\mu)$ is constructed by the light-cone harmonic oscillator model. Then, we obtain $A_1(0)=0.209^{+0.049}_{-0.042}$, $A_2(0)=0.206^{+0.051}_{-0.042}$, and $V(0)=0.258^{+0.058}_{-0.048}$ at the large recoil region. Two important ratios of TFFs are $r_V=1.234_{-0.322}^{+0.425}$ and $r_2=0.985_{-0.274}^{+0.347}$. After extrapolating TFFs to the whole physical $q^2$ region by a simplified $z(q^2,t)$-series expansion, we obtain the differential decay width and branching fraction $\mathcal{B}(B^+\to \omega\ell^+\nu)=(1.35^{+0.02+1.22}_{-0.03-0.66})\times 10^{-4}$, which show good agreement with \textit{BABAR} and Belle Collaborations. Finally, we extract the $|V_{ub}|$ by using the $\mathcal{B}_{\rm{Exp}}(B^+\to \omega\ell^+\nu)$ result from the \textit{BABAR} Collaboration, which leads to $|V_{ub}|=(3.66^{+0.12+1.26}_{-0.17-0.95})\times 10^{-3}$.
\end{abstract}
\maketitle

\section{Introduction}
At present, there are still some gaps between the physical observables detected by many experiments and the predictions of the Standard Model (SM). Many significant observables require an accurate Cabibbo-Kobayashi-Maskawa (CKM) matrix element, which is a unitary $3\times3$ matrix that encodes the probability of a quark flavor transition and is crucial for investigating $CP$ violation beyond the SM. Generally, compared with other rows of CKM matrix, the test of the first row CKM unitarity $|V_{ud}|^2+|V_{us}|^2+|V_{ub}|^2=1$ is more attractive, because it has achieved a very good accuracy in the current experimental and theoretical prediction~\cite{Seng:2018yzq}. Based on the world average results of $|V_{ud}|$, $|V_{us}|$, and $|V_{ub}|$ provided by the 2024 Particle Data Group (PDG)~\cite{ParticleDataGroup:2024cfk}, one can deduce the corresponding accuracies $|\delta V_{\rm{CKM}}|/|V_{\rm{CKM}}|$ as $0.033\%$, $0.38\%$, and $5.2\%$. This reveals that the uncertainty of CKM $|V_{ub}|$ is significantly greater than the other two matrix elements in the first row. Currently, the $|V_{ub}|$ can be determined using the exclusive charmless semileptonic $B$-meson decays, which can describe the decay process more precisely, thereby providing better kinematic constraints and more effective background suppression than inclusive decays, and aiding in more accurately identifying signals in experiments. Examples include $B\to \pi \ell\nu$, $B\to \rho \ell\nu$, $B\to \omega \ell\nu$, etc~\cite{BaBar:2010efp, Belle:2006hlt, CLEO:1999mif, BaBar:2010npl, LHCb:2020ist}. Among them, the world average result of $|V_{ub}|$ is mainly extracted from $B\to \pi \ell\nu$ decay. However, if we study the processes of decay into other mesons, we can investigate observables from another perspective. This approach provides significant benefits for further testing theoretical calculations, improving accuracy, and advancing our understanding of charmless semileptonic decay components. Thus, in this work, we will present a study for the semileptonic $B^+\to \omega \ell^+\nu$ decay, where the $\omega$ is a neutral vector meson.

As early as 1991, while searching for direct evidence of the $b\to u$ transition, the ARGUS Collaboration discovered a decay mode of the $B$ meson in the pattern of $B^+\to \omega \mu^+ \nu$~\cite{ARGUS:1990amj}. This discovery sparked great interest. Immediately after that, in 1993, the CLEO Collaboration reported an upper limit of the branching fraction $\mathcal{B}(B^-\to \omega \ell^- \bar{\nu}) < (1.6-2.7)\times 10^{-4}$~\cite{CLEO:1992adu}. In 2004, the Belle Collaboration directly detected $B^+\to \omega \ell^+\nu$ for the first time in $78~\rm{fb}^{-1}$ of $\Upsilon(4S)$ data accumulated with the Belle detector~\cite{Belle:2004lza}, for which the measured branching fraction was $\mathcal{B}(B^+\to \omega \ell^+\nu)=(1.3\pm0.4\pm0.2\pm0.3)\times 10^{-4}$. Additionally, since 2008, the \textit{BABAR} Collaboration has conducted four measurements of $B^+\to \omega \ell^+\nu$ as the detection accuracy of the \textit{BABAR} detector continuously improved~\cite{BaBar:2008vqc, BaBar:2012dvs, BaBar:2012thb, BaBar:2013pls}. However, the current number of experimental detections for $B^+\to \omega \ell^+\nu$ is still relatively low compared to $B\to \pi\ell\nu$ and $B\to \rho\ell\nu$. Moreover, some discrepancies were observed in the branching fraction measurements when this decay process was reexamined by different experimental collaborations. For example, a new $\mathcal{B}(B^-\to \omega \ell^-\bar{\nu})=(1.07\pm0.16\pm0.07)\times 10^{-4}$ was reported by the Belle Collaboration in 2013~\cite{Belle:2013hlo}, which is significantly smaller than previous measurements. Two separate detections were conducted by the \textit{BABAR} Collaboration in 2012~\cite{BaBar:2012dvs, BaBar:2012thb}, with consistent results. However, in the following year, it updated a branching fraction of $(1.35\pm0.21\pm0.11)\times 10^{-4}$~\cite{BaBar:2013pls}, which showed a significant discrepancy compared to the previous three measurements. This measurement is not only the most recent update from the experimental collaborations but also yields the largest measured result so far. Therefore, based on the existing experimental data, the theoretical predictions for $B^+\to \omega \ell^+\nu$ decay are necessary in order to provide valuable information for determining the parameters of the SM.

Exclusive decay processes involve multiple hadronic matrix elements that must be calculated using physical meson states rather than free quarks. To facilitate calculations, a complete set of Lorentz-invariant transition form factors (TFFs) can be introduced to parametrize these QCD dynamics. The TFFs of $B^+\to \omega$ have been calculated by various methods, such as the QCD light-cone sum rule (LCSR)~\cite{Ball:2004rg,Bharucha:2015bzk}, the heavy quark effective field theory (HQEFT)~\cite{Wu:2006rd}, the soft collinear effective theory (SCET)~\cite{Lu:2007sg}, the perturbative QCD (PQCD) approach~\cite{Lu:2002ny,Li:2009tx}, the light-front quark model (LFQM)~\cite{Verma:2011yw}, and the covariant confined quark model (CCQM)~\cite{Soni:2020bvu}, etc. Among them, the LCSR method has been developed since the 1980s, which can be regarded as an extension of the SVZSR method~\cite{Chernyak:1990ag,Balitsky:1989ry}. At present, it has achieved great success in calculating heavy to light decay processes~\cite{Hu:2024tmc,Aliev:2004vf,Cheng:2017bzz,Fu:2018yin,Ball:2001fp,Ball:2004ye}. Furthermore, it is well known that the LCSR, PQCD, and lattice QCD (LQCD) methods each exhibit distinct advantages in different kinematic $q^2$ ranges, enabling their combined application to provide a more comprehensive description of the TFFs behavior over the entire $q^2$ range. Therefore, utilizing the LCSR method to obtain accurate TFFs is of great importance for conducting systematic research in the future. In this work, we will employ the LCSR method to perform the relevant calculations.

Specifically, a precise understanding of the internal quark structure of mesons constitutes a fundamental requirement for theoretical studies. The neutral mesons $\omega$ and $\rho$ share similar properties as vector meson with spin-parity $J^P=1^-$. However, the constituent parts of a $\omega$ meson are somewhat special in that it may possess a hidden flavor. This distinctive feature arises from its potential to mix with other neutral mesons that also possess hidden flavors and have the same quantum numbers, through both strong and electromagnetic interactions. For example, the $\omega-\phi$ mixing introduces a small $s\bar{s}$ mixing term into the flavor wave function (WF) of a $\omega$ meson, which allows for the $D_s^+\to \omega$ transition. Of course, this decay process may also occur through the weak annihilation contribution. A more detailed discussion can be found in Refs.~\cite{BESIII:2015kin,Gronau:2009mp,Li:2020ylu}. The $B^+\to \omega$ transition discussed in this work is induced by a $b\to u$ transition, and we only need to understand the impact of $\omega-\phi$ mixing on the components of the flavor WF of a $\omega$ meson. At present, there are two main schemes to deal with $\omega-\phi$ mixing. In the framework of singlet-octet mixing scheme, the $\omega$ meson and the $\phi$ meson are mixtures of the SU(3) singlet $\omega_0$ state and octet $\omega_8$ state,
\begin{eqnarray}
\begin{pmatrix} \phi \cr \omega \end{pmatrix}
&=& \begin{pmatrix} \cos\theta_V & -\sin\theta_V \cr
\sin\theta_V &~\cos\theta_V \end{pmatrix} \begin{pmatrix} \omega_8 \cr \omega_0 \end{pmatrix},
\label{mixangle}
\end{eqnarray}
where $\omega_8=(u\bar{u}+d\bar{d}-2s\bar{s})/\sqrt{6}$ and $\omega_0=(u\bar{u}+d\bar{d}+s\bar{s})/\sqrt{3}$. In the quark flavor basis mixing scheme, the physical states of $\omega$ and $\phi$ be further written as
\begin{eqnarray}
|\phi\rangle&=\sin\varphi_V|\omega_q\rangle-\cos\varphi_V|\omega_s\rangle,\\
|\omega\rangle&=\cos\varphi_V|\omega_q\rangle+\sin\varphi_V|\omega_s\rangle.
\label{Eq:FW}
\end{eqnarray}
The quark flavor bases are defined as $|\omega_q\rangle=\frac{1}{\sqrt{2}}(u\bar{u}+d\bar{d})$ and $|\omega_s\rangle=s\bar{s}$. The above equation clearly demonstrates that the mixing angle $\varphi_{V}$ is crucial in determining which flavor bases dominates. For this, the experimental and theoretical studies have been conducted. The mixing angle $\varphi_{V}=(3.4\pm 0.3)^\circ$ was determined by chiral perturbation theory~\cite{Kucukarslan:2006wk}, while a value of $(3.32\pm0.09)^\circ$ was experimentally measured by the KLOE Collaboration~\cite{Ambrosino:2009sc}. A prediction of $\varphi_{V}=3.3^\circ$ was obtained through calculations of $\phi\to\pi^0\gamma$ decays using the chiral SU(3) symmetric Lagrangian~\cite{Klingl:1996by}. Related discussion can also be found in Refs.~\cite{Benayoun:2009im,Benayoun:2007cu,Choi:2015ywa}. Current theoretical predictions and experimental measurements of the mixing angle $\varphi_{V}$ show certain discrepancies, while all results consistently satisfy $\cos\varphi_{V}\approx 1$ within their respective uncertainties. As can be seen from Eq.~\eqref{Eq:FW}, the first term dominates. Therefore, in this paper, the neutral vector meson $\omega$ is considered as a pure meson that only contains $u$ and $d$ quarks.

In LCSR approach, the operator product expansion (OPE) is carried out near the light-cone $x^2\rightsquigarrow 0$. The nonperturbative effects are replaced by physically meaningful hadronic light-cone distribution amplitudes (LCDAs) with the progressively increasing twists. It is worth noting that the LCDAs of a vector meson $\omega$ comprise two distinct polarization states, which arise from the chiral-odd and chiral-even operators. We denote the longitudinal part by the symbol ``$\|$'' and the transverse part by the symbol ``$\perp$'' in this paper. Meanwhile, a process-dependent kinematic parameter $\delta$ is introduced to classify the contributions of different LCDAs to TFFs~\cite{Ball:2004rg,Ball:1998sk}. For a vector meson $\omega$, $\delta\simeq m_\omega/m_b\sim 16\%$.  Up to the $\delta^3$ order, the $\omega$ meson has fifteen LCDAs. To avoid the computational complexity induced by considering all LCDAs, we will adopt an improved LCSR method. This method involves selecting chiral currents to replace traditional currents when constructing the correlation function~\cite{Fu:2014pba,Wang:2002aje,Huang:2001xb,Huang:1998gp}, which can achieve the effect of highlighting the contributions from LCDA that we primarily focus on. In this work, we will employ left-handed chiral current to directly or indirectly make twist-2 LCDA $\phi^\| _{2;\omega}(x,\mu)$ contributions dominant.

Generally, the vector meson's LCDA $\phi^\| _{2;\omega}(x,\mu)$ can be expanded with a series of Gegenbauer coefficients, and it is common to adopt a truncated form retaining only the first few terms~\cite{Hua:2020gnw,LatticeParton:2022zqc}. Meanwhile, the behavior of $\phi^\| _{2;\omega}(x,\mu)$ can also be described through other phenomenological models, which can offer a new perspective to assist us in deepening our understanding of the meson structure. In this work, we will adopt the light-cone harmonic oscillator (LCHO) model to construct a $\omega$ meson twist-2 LCDA $\phi^\| _{2;\omega}(x,\mu)$, which is based on the Brodsky-Huang-Lepage (BHL) description that suggests the hadronic WF can be determined by connecting the equal-time WF in the rest frame and the WF in the infinite momentum frame~\cite{Zhong:2011rg, Wu:2007rt, Huang:2004su}. This model has been successfully applied to the pseudoscalar mesons $\pi, K, \eta^{(\prime )}$~\cite{Zhong:2022ecl, Zhong:2021epq, Hu:2023pdl}, scalar mesons $a_0(980), K_0^*(1430)$~\cite{Huang:2022xny,Yang:2024ang,Wu:2022qqx}, and vector mesons $\rho, K^*, \phi$~\cite{Fu:2014uea,Fu:2016yzx,Hu:2024tmc}.

This paper is organized as follows. In Sec.~\ref{Sec:2}, we present the calculation for the TFFs of $B^+\to \omega$ within the LCSR method and construct the twist-2 LCDA $\phi^\| _{2;\omega}(x,\mu)$ by LCHO model. In Sec.~\ref{Sec:3}, we show the detailed numerical analysis for TTFs, differential decay width, branching fraction and CKM matrix element $|V_{ub}|$. The Sec.~\ref{Sec:4} is used to be a brief summary.
\section{Theoretical framework}\label{Sec:2}
The effective Hamiltonian for semileptonic $b\to u \ell^+\nu$ from the four-Fermi interaction in SM can be written as
\begin{align}
\mathcal{H}^{\rm{SM}}_{\rm{eff}}=\frac{4G_{F}}{\sqrt{2}}V_{ub}(\bar{u}\gamma_{\mu}P_{L}b)(\bar{\nu}\gamma^{\mu}P_{L}\ell^+),
\label{Eq:Heff}
\end{align}
where the Fermi coupling constant $G_{F}=1.166\times 10^{-5}~{\rm GeV}^{-2}$ and the chiral projects $P_{L}=(1-\gamma_{5})/2$. In order to obtain the free quark amplitude $\mathcal{M}(B^+\to \omega\ell^+\nu)$ explicitly, we need to sandwich Eq.~\eqref{Eq:Heff} between the initial and final meson states, $i.e.,$ $\langle \omega(p,\lambda)|\bar{u}\gamma_{\mu}(1-\gamma_5)b|B^+(p+q) \rangle$. And this matrix element must be calculated between physical final hadronic states that contain nonperturbative strong interaction contributions. These difficult-to-calculate quantities can be parametrized in terms of the Lorentz-invariant TFFs as follows:
\begin{align}
&\langle \omega(p,\lambda)|\bar{u}\gamma_{\mu}(1-\gamma_5)b|B^+(p+q) \rangle\nonumber\\
&\qquad\qquad =  -i e_{\mu}^{*(\lambda)}(m_{B^+}+m_\omega) A_{1}(q^2) \nonumber\\
&\qquad\qquad +i(e^{*(\lambda)} \cdot q) \frac{A_{2}(q^2)(2 p+q)_{\mu}}{m_{B^+}+m_\omega}\nonumber\\
&\qquad\qquad +i q_{\mu}(e^{*(\lambda)} \cdot q) \frac{2 m_\omega}{q^2 }[A_{3}(q^2)-A_{0}(q^2)]\nonumber\\
&\qquad\qquad +\epsilon_{\mu \nu \alpha \beta} e^{*(\lambda) \nu} q^{\alpha} p^{\beta} \frac{2 V(q^2)}{m_{B^+}+m_\omega},
\label{Eq:matrix difen}
\end{align}
where $m_{B^+}$, $m_\omega$, $p$, and $q=(p_{B^+}-p_\omega)$ are the masses of the $B^+$ meson and $\omega$ meson, $\omega$ meson momentum, and the four-momentum transfer between those two mesons, respectively. The $e^{*(\lambda)}$ stands for the $\omega$ meson polarization vector with $\lambda$ being its transverse $(\perp)$ or longitudinal $(\|)$ polarization. In addition, the $A_{0,1,2}(q^2)$ and $V(q^2)$ are four semileptonic TFFs, and $A_3(q^2)$ is not independent which satisfies the following relationship:
\begin{align}
A_{3}(q^2)=\frac{m_{B^+}+m_\omega}{2 m_\omega} A_{1}(q^2)-\frac{m_{B^+}-m_\omega}{2 m_\omega} A_{2}(q^2) .
\end{align}

In the SM, the differential decay width after neglecting the masses of charged lepton can be written as~\cite{Gao:2019lta,Gilani:2003hf}
\begin{align}
\frac{d\Gamma (B^+\to \omega\ell^+ \nu )}{dq^2d\cos \theta _{W\ell}}&=|V_{ub}|^2 \frac{G_{F}^2 |\boldsymbol{p}_\omega|}{128\pi ^3m_{B^+}^2 }\frac{q^2}{c_\omega^2}
\nonumber\\
&\times \bigg[ (1-\cos \theta _{W\ell})^2\frac{|H_+(q^2)|^2}{2}
\nonumber\\
& +(1+\cos \theta _{W\ell})^2\frac{|H_-(q^2)|^2}{2}
\nonumber\\
&+\sin ^2\theta _{W\ell}|H_0(q^2)|^2 \bigg],
\label{Eq:dGammacos}
\end{align}
in which the $\theta _{W\ell}$ is the angle between the direction of the charged lepton in the virtual $W$-gauge boson rest frame and the direction of the virtual $W$ in the $B^+$-meson rest frame. The isospin factor $c_\omega$ is equal to $\sqrt{2}$ for $B^+\to \omega\ell^+\nu$~\cite{Ball:2004rg}. And the three-momentum $|\boldsymbol{p}_\omega|$ of $\omega$ meson has the following form:
\begin{align}
&|\boldsymbol{p}_\omega|=\frac{1}{2m_{B^+}}\lambda^{1/2}(m_{B^+}^2,m_\omega^2,q^2), \nonumber\\
&\lambda(a,b,c)=a^2+b^2+c^2-2ab-2ac-2bc.
\end{align}
Moreover, in the standard helicity basis, the three helicity amplitudes $H_i(q^2)$ with $i=(\pm,0)$ can in turn be related to the axial-vector TFFs $A_1(q^2)$ and $A_2(q^2)$, and the vector TFF $V(q^2)$, $i.e.,$
\begin{align}
H_{\pm}(q^2)&=(m_{B^+}+m_\omega)\Big[ A_1(q^2)\mp \frac{2m_{B^+}|\boldsymbol{p}_\omega|}{(m_{B^+}+m_V)^2}\nonumber\\
&\times V(q^2) \Big],\nonumber\\
	H_0(q^2)&=\frac{m_{B^+}+m_\omega}{2m_\omega\sqrt{q^2}}\bigg[ (m_{B^+}^2 -m_\omega^2 -q^2)A_1(q^2)\nonumber\\
&-\frac{4m_{B^+}^2 |\boldsymbol{p}_\omega|^2}{(m_{B^+}+m_\omega)^2}A_2(q^2) \bigg].
\end{align}
From the equations above, we can see that $A_2(q^2)$ contributes only to $H_0(q^2)$, $V(q^2)$ contributes only to $H_{\pm}(q^2)$, while $A_1(q^2)$ contributes to all helicity amplitudes $H_{\pm,0}(q^2)$. In the high $q^2$ region, the $A_1(q^2)$ will dominate the contributions. After integrating Eq.~\eqref{Eq:dGammacos} over the angle $\cos \theta _{W\ell}$, we can obtain
\begin{align}
&\frac{d\Gamma (B^+\rightarrow \omega \ell ^+\nu)}{dq^2}=
|V_{ub}|^2\frac{G_{F}^2 q^2|\boldsymbol{p}_\omega|}{96\pi^3 m_{B^+}^2 c_\omega^2 }\nonumber\\
&~~~~~~~~~~~\times \bigg[ |H_0(q^2)|^2+|H_+(q^2)|^2+|H_-(q^2)|^2 \bigg] .
\label{Eq:dGamma}
\end{align}
According to polarization states of these helicity amplitudes, the differential decay width can be decomposed into longitudinal polarization state and transverse polarization state, $i.e.$,
\begin{align}
\frac{d\Gamma_{\rm{L}} (B^+\rightarrow \omega \ell ^+\nu)}{dq^2}&=
|V_{ub}|^2\frac{G_{F}^2 q^2|\boldsymbol{p}_\omega|}{96\pi^3 m_{B^+}^2 c_\omega^2 } |H_0(q^2)|^2  .
\label{Eq:dGammaL}
\end{align}
and
\begin{align}
\frac{d\Gamma_{\pm} (B^+\rightarrow \omega \ell ^+\nu)}{dq^2}&=
|V_{ub}|^2\frac{G_{F}^2 q^2|\boldsymbol{p}_\omega|}{96\pi^3 m_{B^+}^2 c_{\omega}^2 } |H_{\pm}(q^2)|^2  .
\label{Eq:dGammaT}
\end{align}
In which the specific transverse polarization part is $d\Gamma_{\rm{T}}=d\Gamma_{+}+d\Gamma_{-}$. By integrating over $q^2$ in entire physical region and utilizing the lifetime $\tau_{B^+}$, the branching fraction can be obtained, $i.e.,$
\begin{align}
\mathcal{B}(B^+\rightarrow \omega \ell ^+\nu)&=\tau_{B^+} \int_0^{q^2_{\rm{max}}}\frac{d\Gamma (B^+\rightarrow \omega \ell ^+\nu)}{dq^2} dq^2
\label{Eq:Br}
\end{align}
with $q^2_{\rm{max}}=(m_{B^+}-m_{\omega})^2$. The three TFFs serve as the primary nonperturbative input parameters, and accurately determining their behavior is crucial for measuring the branching fraction and extracting the CKM matrix element $|V_{ub}|$.

For the next step, we can start from the vacuum to meson correlation function to derive the LCSR of the TFFs firstly. The detailed form can be written as
\begin{align}
\Pi _{\mu}(p,q)&=i\int d^4xe^{iq\cdot x}\langle \omega (p,\lambda )|\mathrm{T}\{ \bar{q}_1(x)\gamma _{\mu}(1-\gamma _5)b(x),
\nonumber\\
&\times j_{B^+}^{\dagger}(0)\} |0\rangle.
\label{Eq:correlation function}
\end{align}
For $j_{B^+}^{\dagger}(x)$, we choose the left-handed chiral current $i\bar{b}(x)(1-\gamma _5)q_2(x)$ instead of the traditional current $i\bar{b}(x)\gamma _5q_2(x)$. This method can eliminate the contributions from the chiral-odd LCDAs, ensuring that the nonperturbative inputs in TFFs are primarily determined by the chiral-even LCDAs. Following to the basic steps of the QCDSR, the correlation function, Eq.~\eqref{Eq:correlation function}, can be inserted a complete set of states with the same quantum numbers as $B^+$ meson in a timelike $q^2$ region. Here, the matrix element $\langle B^+(p+q)|\bar{b}i\gamma _5q_2|0\rangle =m_{B^+}^2 f_{B^+}/m_b$ for the decay constant of $B^+$ meson and Eq.~\eqref{Eq:matrix difen} will be used. After separating the pole term of the lowest pseudoscalar $B^+$ meson and replacing the contributions from higher resonances and continuum states with a dispersion relation, we can derive the hadronic representation of the correlation function. On the other hand, in a spacelike $q^2$ region, the Eq.~\eqref{Eq:correlation function} can be calculated by QCD theory. To be specific, the OPE can be carried out near the light-cone $x^2\rightsquigarrow 0$, which corresponds to $(p+q)^2-m_b^2\ll0$ with the momentum transfer $q^2\sim\mathcal{O}(1~{\rm GeV}^2) \ll$$m_b^2$ and ensures the validity of OPE. Finally, by utilizing the Borel transform to suppress contributions from higher twists and continuum states, and employing quark-hadron duality to match the OPE result with hadronic representation, we can obtain the analytic expression for the TFFs within the framework of LCSR. The corresponding result is similar to our previous work on $B\to\rho$~\cite{Fu:2014cna}, which has been verified through our recalculation. Then, under the LCSR method, the final analytic expression of TFFs can be written as follows:
\begin{widetext}
\begin{align}
A_1(q^2)&=\frac{2m_{b}^2 m_\omega f_\omega ^\| e^{m_{B^+}^2 /M^2}}{m_{B^+}^2 (m_{B^+}+\! m_\omega )f_{B^+}}
\bigg\{ \int_{u_0}^1 \! du e^{-s(u)/M^2} \!\left[ \frac{1}{u}\phi _{3;\omega}^{\bot}(u) \!-\! \frac{m_\omega ^2 }{u^2M^2}C_\omega ^\| (u)\right] \!-\! e^{-s_0/M^2}\! \frac{m_\omega ^2 }{m_{b}^2 + \! u_{0}^2 m_\omega ^2 -q^2}
\nonumber\\
&\times C_\omega ^\| (u_0)
-m_\omega ^2 \int{\mathcal{D} \underline{\alpha }\int{dve^{-s(X)/M^2}\frac{1}{X^2M^2}\Theta (c(X,s_0))[\Phi _{3;\omega}^\| (\underline{\alpha })}}+\widetilde{\Phi }_{3;\omega}^\| (\underline{\alpha })] \bigg\},
\label{Eq:A1TFFs}
\\
A_2(q^2)&=\frac{m_{b}^2 \, m_\omega (m_{B^+}+m_\omega )f_\omega ^\| e^{m_{B^+}^2 /M^2}}{m_{B^+}^2 f_{B^+}} \, \biggl\{ 2\int_{u_0}^1 du \, e^{-s(u)/M^2}\bigg[\frac{1}{u^2M^2}\, A_\omega ^\| (u)+ \frac{m_\omega ^2 }{u^2M^4}\, C_\omega ^\| (u)+\frac{m_{b}^2 m_\omega ^2 }{4u^4M^6}
\nonumber\\
&\times \, B_\omega ^\| \, (u)\,  \bigg]~ +~ 2 e^{-s_0/M^2}\, \bigg[ \, \frac{A_\omega ^\| (u_0)}{m_{b}^2 \,
+ \, u_{0}^2 m_\omega ^2 \, - \,  q^2}~ +~ \frac{m_\omega ^2 }{M^2} \frac{C_\omega ^\| (u_0)}{m_{b}^2 \, +\, u_{0}^2 m_\omega ^2 \, -\, q^2} ~ - ~ \frac{m_\omega ^2  u_0^3}
{m_{b}^2 \, +\, u_{0}^2 m_\omega ^2 \, -\, q^2}
\nonumber\\
& \times\frac{d}{du}\, \bigg(\frac{C_\omega ^\| (u)}{u(m_{b}^2 \, +u^2 m_\omega ^2 \, -q^2)}\bigg)\bigg|_{u=u_0}\, \bigg] \, +\, m_\omega ^2 \int{\mathcal{D}\underline{\alpha}\, \int{dv e^{-s(X)/M^2}}}\, \frac{1}{X^3M^4}\, \Theta (c(X,s_0))\, [ \Phi _{3;\omega}^\| (\underline{\widetilde{\alpha }})
\nonumber\\
&+\widetilde{\Phi }_{3;\omega}^\| (\underline{\widetilde{\alpha }})]\bigg\},
\label{Eq:A2TFFs}
\\
V(q^2)&=\frac{m_{b}^2 m_\omega (m_{B^+}+m_\omega )f_\omega ^\| e^{m_{B^+}^2 /M^2}}{2m_{B^+}^2 f_{B^+}}
\bigg[\, \int_{u_0}^1due^{-s(u)/M^2}\frac{1}{u^2M^2}\psi _{3;\omega}^{\bot}(u)+e^{-s_0/M^2}\frac{\psi _{3;\omega}^{\bot}(u_0)}{m_{b}^2 +u_{0}^2 m_\omega ^2 -q^2}\, \bigg],
\label{Eq:VTFFs}
\end{align}
\end{widetext}
with
\begin{align}
u_0&=\bigg[ \sqrt{(q^2-s_0+m_\omega ^2)^2+4 m_\omega ^2(m_b^2-q^2)} \nonumber\\
&+q^2-s_0+m_\omega ^2 \bigg]/(2m_\omega ^2),
\end{align}
where $s(u)=[m_{b}^2 -\bar{u}(q^2-u m_\omega ^2 )]/u$ with $\bar{u}=1-u$, $X=a_1+va_3$, and $c\left( u,s_0 \right) =us_0-m_{b}^2 +\overline{u}q^2-u\overline{u}m_\omega ^2 $. $\Theta (c(X ,s_0))$ is the usual step function. For two-particle twist-3 LCDAs $\psi^{\bot}_{3;\omega}(x,\mu)$ and $\phi^{\bot}_{3;\omega}(x,\mu)$, due to the existence of polarization states of the vector mesons and the non-negligible mass of the meson, the expression is more complex compared to that of the pseudoscalar meson, making the treatment process more challenging. To address this, Ball and Braun presented a systematic study of twist-3 LCDA for vector mesons in 1998~\cite{Ball:1998sk}. Based on conformal symmetry and QCD equations of motion, they demonstrated that the $\psi^{\bot}_{3;\omega}(x,\mu)$ and $\phi^{\bot}_{3;\omega}(x,\mu)$ can be eliminated in favor of independent dynamical degrees of freedom, transforming into the leading twist LCDA, three-particle LCDA, and quark mass correction terms.
\begin{align}
&\hspace{-0.2cm}\phi _{3;\omega}^{\bot}(x,\mu )=\phi _{3;\omega}^{\bot WW}(x,\mu )\!+\!\phi _{3;\omega}^{\bot g}(x,\mu )\!+\!\phi _{3;\omega}^{\bot m}(x,\mu ),\nonumber\\
&\hspace{-0.2cm}\psi _{3;\omega}^{\bot}(x,\mu )=\psi _{3;\omega}^{\bot WW}(x,\mu )\!+\!\psi _{3;\omega}^{\bot g}(x,\mu )\!+\!\psi _{3;\omega}^{\bot m}(x,\mu ),
\end{align}
where contribution from the three-particle LCDA $\phi _{3;\omega}^{\bot g}(x,\mu )$ is particularly small and can reasonably be ignored. The quark mass correction terms $\phi _{3;\omega}^{\bot m}(x,\mu )$ has a coefficient $\widetilde{\delta}_{\pm}$, which is proportional to the quark mass in the meson components system. For a $\omega$ meson, this coefficient tends to zero~\cite{Ball:1998sk}. Then, by using the Wandzura-Wilczek (WW) approximation, the twist-3 LCDAs $\psi^{\bot}_{3;\omega}(x,\mu)$ and $\phi^{\bot}_{3;\omega}(x,\mu)$ can be defined as
\begin{align}
&\phi_{3;\omega}^{\bot}(x,\mu)\!=\!\frac{1}{2}\left[ \int_0^{x}{dv\frac{\phi _{2;\omega}^\| (v,\mu )}{\bar{v}}}+\int_x^{1}{dv\frac{\phi _{2;\omega}^\| (v,\mu )}{v}} \right]\!,\nonumber\\
&\psi_{3;\omega}^{\bot}(x,\mu)\!=\!2\!\left[\! \bar{x}\!\int_0^{x}dv\frac{\phi _{2;\omega}^\| (v,\!\mu )}{\bar{v}}\!+\!x\int_x^{1}\!dv\frac{\phi _{2;\omega}^\| (v,\!\mu )}{v}\right]\!.
\label{Eq:WWA}
\end{align}
And the simplified function $A^\| _\omega (u)=\int_0^udv[\phi_{2;\omega}^\| (v)-\phi^{\bot}_{3;\omega}(v)]$ can be further written as
\begin{align}
A_\omega ^\| (x,\mu )&=\frac{1}{2}\left[ \bar{x}\int_0^{x}{dv\frac{\phi _{2;\omega}^\| (v,\mu )}{\bar{v}}}+x\int_x^{1}{dv\frac{\phi _{2;\omega}^\| (v,\mu )}{v}} \right].
\label{Eq:WWA1}
\end{align}
The remaining simplified LCDA is
\begin{align}
C^\| _\omega (u)=\int^u_0dv\int^v_0dw[\psi^\| _{4;\omega}+\phi^\| _{2;\omega}-2\psi^{\bot}_{3;\omega}(w)],
\end{align}
which can be found in our previous work~\cite{Fu:2014cna}.

Moreover, we need to construct the twist-2 LCDA $\phi _{2;\omega}^\| (x,\mu)$ by using LCHO model. The relationship between twist-2 LCDA of $\omega$ meson and WF is defined as
\begin{align}
\phi _{2;\omega}^\| (x,\mu)=\frac{2\sqrt{3}}{\widetilde{f}_\omega ^\| }\int_{|\mathbf{k}_{\bot}|^2\le \mu _{0}^2 }{\frac{d^2\mathbf{k}_{\bot}}{16\pi ^3}\psi _{2;\omega}^\| (x,\mathbf{k}_{\bot})},
\end{align}
where $\widetilde{f}_\omega ^\| =f_\omega ^\| /\sqrt{5}$ and $\mathbf{k}_{\bot}$ is the $\omega$ meson transverse momentum. Based on the BHL description~\cite{Zhong:2011rg,Wu:2007rt,Huang:2004su}, the light meson WF $\psi _{2;\omega}^\| (x,\mathbf{k}_{\bot})$ will include the spin WF $\chi _{2;\omega}(x,\mathbf{k}_{\bot})$ and spatial WF $\Psi^{R}_{2;\omega}(x,\mathbf{k}_{\bot})$, $i.e.,$
\begin{align}
\psi _{2;\omega}^\| (x,\mathbf{k}_{\bot})=\chi _{\lambda_1,\lambda_2}^{L}(x,\mathbf{k}_{\bot})\Psi^{R}_{2;\omega}(x,\mathbf{k}_{\bot}),
\end{align}
in which $L$ indicates the longitudinal spin projection for $\omega$ meson. $\lambda_1$ and $\lambda_2$ are the different helicities of quark and antiquark. $\chi _{\lambda_1,\lambda_2}^{L}(x,\mathbf{k}_{\bot})$ can be determined by the Wigner-Melosh rotation, which establishes a connection between the spin states transforming from the instant form to the light-front form to address the spin structure of the WF. Based on this, the longitudinal and transverse spin form of $\rho$ meson is derived by light-front holographic model~\cite{Kaur:2020emh}. Here, since the $\omega$ meson and the $\rho$ meson possess the same spin projection and this work focuses on the longitudinal twist-2 LCDA, using the same treatment, we can consider the following form of the spin WF:
\begin{align}
\chi _{2;\omega}(x,\mathbf{k}_{\bot})=\frac{\hat{m}_q(\mathcal{M}+2\hat{m}_q)+2\mathbf{k}_{\bot}^2}{(\mathcal{M}+2\hat{m}_q)\sqrt{2(\mathbf{k}_{\bot}^2+\hat{m}_q)}}.
\end{align}
with $\mathcal{M}=\sqrt{(\mathbf{k}_{\bot}^2+\hat{m}_q^2)/(x\bar{x})}$. Additionally, $\Psi^{R}_{2;\omega}(x,\mathbf{k}_{\bot})$ can be separated into two components: the $x$-dependent part $\varphi(x)$ and the $\mathbf{k}_{\bot}$-dependent part. The latter arising from the harmonic oscillator solution for the meson in its rest frame,
\begin{align}
\Psi^{R}_{2;\omega}(x,\mathbf{k}_{\bot})&= A_{2;\omega}^\| \varphi(x)
\exp \left( -b_{2;\omega}^{\|2} \frac{\mathbf{k}_{\bot}^2 +\hat{m}_q^2}{x\bar{x}}\right),
\end{align}
where $A^\| _{2;\omega}$ is the normalization constant, $b^\| _{2;\omega}$ is the harmonic parameter, and $\hat{m}_q\simeq300~\rm{MeV}$ is constituent quark mass. The function $\varphi(x)=1+B^\| _{2;\omega}C^{3/2}_2 (2x-1)$ with Gegenbauer polynomials $C^{3/2}_2(2x-1)$. Among them, $B^\| _{2;\omega}$ dominates the longitudinal distribution. Then, the twist-2 LCDA $\phi _{2;\omega}^\| (x,\mu)$ can be written as
\begin{align}
\phi _{2;\omega}^\| (x,\mu)&=\frac{2\sqrt{3}}{\widetilde{f}_\omega ^\| }\int_{|\mathbf{k}_{\bot}|^2\le \mu _{0}^2 }{\frac{d^2\mathbf{k}_{\bot}}{16\pi ^3}}A^\| _{2;\omega}\varphi(x)\nonumber\\
&\times \frac{\hat{m}_q (\mathcal{M}+2\hat{m}_q)+2\mathbf{k}_{\bot}^2} {(\mathcal{M}+2\hat{m}_q)\sqrt{2(\mathbf{k}_{\bot}^2+\hat{m}_q)}}\nonumber\\
&\times \exp \left[ -b_{2;\omega}^{\|2} \frac{\mathbf{k}_{\bot}^2 +\hat{m}_q^2}{x\bar{x}}\right].
\end{align}
The remaining three unknown parameters, $A^\| _{2;\omega}$, $b_{2;\omega}^\| $, and $B^\| _{2;\omega}$  need to be determined using additional conditions:
\begin{itemize}
\item The normalization condition of $\omega$ meson twist-2 LCDA,
\begin{align}
\int_0^1dx \phi _{2;\omega}^\| (x,\mu)=1.
\end{align}
\item The average value of the squared transverse momentum $\langle \mathbf{k}_{\bot}^2 \rangle_{2;\omega}$,
\begin{align}
\langle \mathbf{k}_{\bot}^2 \rangle _{2;\omega}=\frac{\int{dxd^2\mathbf{k}_{\bot}|\mathbf{k}_{\bot}|^2|\psi _{2;\omega}^\| (x,\mathbf{k}_{\bot})|^2}}{\int{dxd^2\mathbf{k}_{\bot}|\psi _{2;\omega}^{\|
}(x,\mathbf{k}_{\bot})|^2}}.
\end{align}
Here we use $\langle \mathbf{k}_{\bot}^2 \rangle _{2;\omega}^{1/2}=0.37~{\rm GeV}$~\cite{Guo:1991eb} to do numerical calculation.
\item The Gegenbauer moments $a_{n;\omega}^\| (\mu)$ can be derived by the following way:
\begin{align}
a_{n;\omega}^\| (\mu)=\frac{\int_0^1{dx\phi _{2;\omega}^\| (x,\mu)C_{n}^{3/2}(\xi)}}{\int_0^1{6x\bar{x}[C_{n}^{3/2}(\xi)]^2}},
\label{Eq:a2}
\end{align}
with $\xi=(2x-1)$. Here, we adopt $a_{2;\omega}^\| (\mu_0)=0.15\pm0.12$ determined by Dimoul and Lyon (DL) at an initial scale $\mu_0=1~\rm{GeV}$~\cite{Dimou:2012un}, which was obtained by considering the value of $a_{2;\rho}^\| (\mu_0)$ from both RBC and UKQCD Collaborations~\cite{Arthur:2010xf}, and QCDSR~\cite{Ball:2007rt}, with the original uncertainties conservatively doubled.
\end{itemize}

\section{NUMERICAL ANALYSIS}\label{Sec:3}
\begin{figure}[t]
\begin{center}
\includegraphics[width=0.435\textwidth]{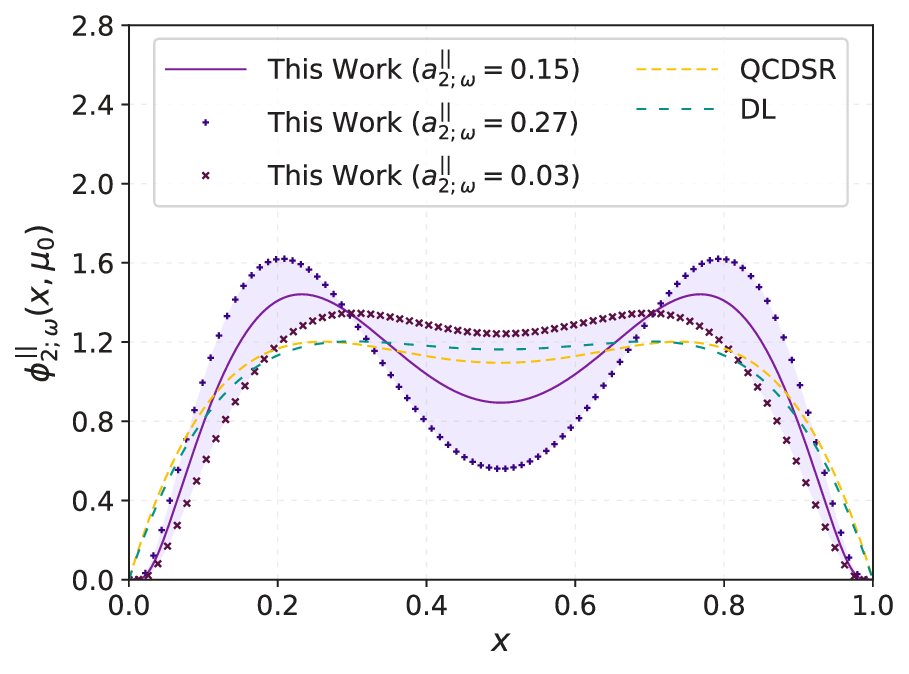}
\caption{The behavior of $\phi _{2;\omega}^\| (x,\mu_0)$ in LCHO model within uncertainties. The QCDSR~\cite{Ball:1998sk} and DL~\cite{Dimou:2012un} are also used for comparison.}
\label{Fig:DA}
\end{center}
\end{figure}

\begin{figure}[h]
\begin{center}
\includegraphics[width=0.435\textwidth]{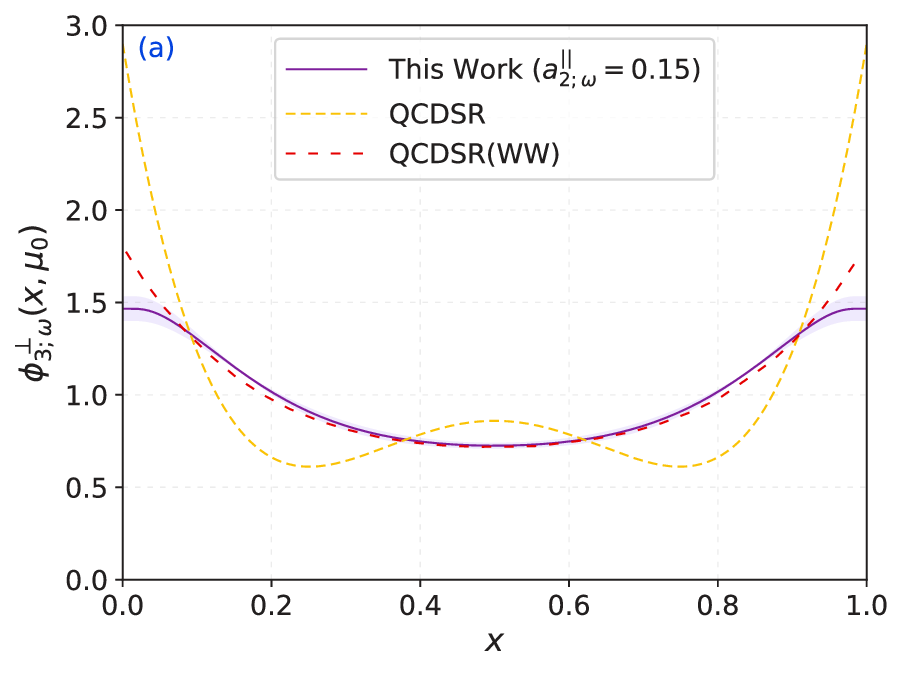}
\includegraphics[width=0.435\textwidth]{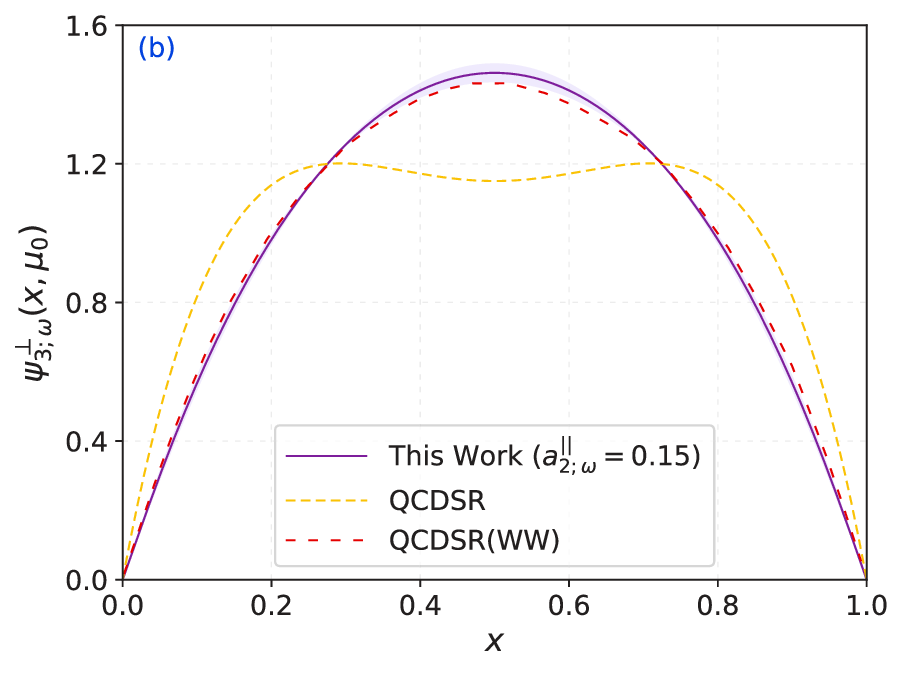}
\caption{The behavior of (a) $\phi _{3;\omega}^{\bot}(x,\mu_0)$ and (b) $\psi _{3;\omega}^{\bot}(x,\mu_0)$ under the WW approximation, which encompasses both the asymptotic distributions of QCDSR~\cite{Ball:1998sk} and the distributions under the WW approximation.}
\label{Fig:twist3DA}
\end{center}
\end{figure}

Before performing numerical calculations, we need to define some basic input parameters from PDG~\cite{ParticleDataGroup:2024cfk}: $m_{B^+}=5279.41\pm0.07~\rm{MeV}$, $m_\omega =782.66\pm0.13~\rm{MeV}$, and the pole mass of $b$ quark $m_b= 4.78\pm0.06~{\rm GeV}$; the decay constant $f_{B^+}=0.160\pm0.019~{\rm GeV}$~\cite{Fu:2014pba} and $f_\omega =0.197\pm0.008~{\rm GeV}$~\cite{Bharucha:2015bzk}. For $B^+\to \omega$ transition, we take the typical process energy scale $\mu_k=(m_{B^+}^2-m_b^2)^{1/2}\simeq 2.2~{\rm GeV}$. As demonstrated in Eqs.~\eqref{Eq:WWA}-\eqref{Eq:WWA1}, the twist-2 LCDA $\phi _{2;\omega}^\| (x,\mu)$ provides the dominant contribution either directly or indirectly to the TFFs. Therefore, we need to first determine the behavior of $\phi _{2;\omega}^\| (x,\mu)$. By utilizing the three conditions elaborated in the previous section, namely normalization, the average value of squared transverse momentum, and Gegenbauer moment, we are enabled to determine the unknown parameters within $\phi _{2;\omega}^\| (x,\mu)$. Meanwhile, in order to obtain the LCDA parameters at a corresponding process energy scale , $a_{2\omega}^\| (\mu)$ needs to evolve from the initial energy scale $\mu_0$ to $\mu_k$ through the renormalization group equation,
\begin{align}
a_{n;\omega}^\| (\mu_k)=a_{n;\omega}^\| (\mu_0)\left(\frac{\alpha_s(\mu_0)}{\alpha_s(\mu_k)}\right)^{\frac{-\gamma_n+4}{b}},
\end{align}
with $b=(33-2n_f)/3$ and the one-loop anomalous dimensions~\cite{Braun:2003rp},
\begin{align}
\gamma_n=C_F\left( 1-\frac{2}{(n+1)(n+2)}+4\sum^{n+1}_{j=2}\frac{1}{j}\right)
\end{align}
Then, the detailed numerical values can be determined and listed in Table~\ref{table:modelparameter}, which also includes the parameters corresponding to the upper and lower limits of $a_{2;\omega}^\| (\mu)$ at different scales $\mu$. Additionally, we provide the predicted results for $a^\| _{4;\omega}(\mu_0)$ through Eq.~\eqref{Eq:a2}, $i.e.$,
\begin{align}
a^\| _{4;\omega}(\mu_0)=-0.12\pm0.03.
\end{align}
Subsequently, the specific behavior of $\phi _{2;\omega}^\| (x,\mu_0)$ can be determined and is presented in Fig.~\ref{Fig:DA}, where the error band arises from the uncertainty in $a_{2;\omega}^\| (\mu_0)$. For ease of comparison, we represent the central result with a solid line and also encompass the QCDSR~\cite{Ball:1998sk} and the DL~\cite{Dimou:2012un} based on the truncated form of Gegenbauer coefficients. Notably, our central result exhibits more pronounced double-peak behavior compared to the results of QCDSR and DL. Meanwhile, the asymptotic distributions and the distributions under the WW approximation for the twist-3 LCDAs $\phi _{3;\omega}^{\bot}(x,\mu_0)$ and $\psi _{3;\omega}^{\bot}(x,\mu_0)$ are also presented in QCDSR~\cite{Ball:1998sk}. The latter will be referred to as QCDSR(WW) in our paper. We present their results in Fig.~\ref{Fig:twist3DA} and compare them with ours. It is clearly visible that after applying the WW approximation to these two LCDAs, the QCDSR(WW) is very close to our predicted results, and both exhibit a single-peak behavior. This also demonstrates that the LCHO model of a $\omega$ meson possesses good feasibility and is capable of providing us with a novel perspective for describing the internal structure and dynamics of mesons.
\begin{table}
\begin{center}
\renewcommand{\arraystretch}{1.5}
\footnotesize
\caption{The three model parameters of $\omega$ meson twist-2 LCDA $\phi _{2;\omega}^\| (x,\mu)$, $A_{2;\omega}^\| ~({\rm GeV}^{-1})$, $b_{2,\omega}^\| ~({\rm GeV}^{-1})$, and $B_{2;\omega}^\| $. The obtained results correspond to the upper limit, central value, and lower limit of $a_{2;\omega}^\| (\mu)$ at $\mu_0$ and $\mu_k$, respectively.}
\label{table:modelparameter}
\begin{tabular}{l l l l}
\hline
       ~~~~~~&$A_{2;\omega}^\| $  ~~~~~~   &$b_{2;\omega}^\| $    ~~~~~~ &$B_{2;\omega}^\| $ ~~~~~\\
\hline
                     &122.77                   &0.83                    &0.47      \\           \cline{2-4}
$\mu_0=1.0$~{\rm GeV} ~~~            &135.91                   &0.87                    &0.35   \\            \cline{2-4}
                     &151.64                   &0.92                    &0.22    \\             \cline{2-4}
\hline
                     &129.17                  &0.86                    &0.38  \\                \cline{2-4}
$\mu_k=2.2$~{\rm GeV}  ~~~            &138.81                   &0.89                    &0.30   \\          \cline{2-4}
                     &148.95                   &0.92                    &0.22    \\             \cline{2-4}
\hline
\end{tabular}
\end{center}
\end{table}

Furthermore, the calculation of $B^+\to \omega$ TFFs requires two crucial input parameters: the continuum threshold $s_0$ and Borel parameter $M^2$. Typically, the continuum threshold $s_0$ is set near the squared mass of $B^+$ meson's first excited state. Secondly, we require that the contribution from the higher excited resonances and continuum states does not exceed $30\%$ in the total sum rules, and the contribution from high twist does not exceed $5\%$. Then we have
\begin{align}
&s^{A_1}_0= 35\pm1~{\rm GeV}^2,    &&M^2_{A_1}=2.75\pm0.10~{\rm GeV}^2.\nonumber\\
&s^{A_2}_0= 36\pm1~{\rm GeV}^2,    &&M^2_{A_2}=2.75\pm0.10~{\rm GeV}^2.\nonumber\\
&~s^{V}_0= 35\pm1~{\rm GeV}^2,   ~&&~M^2_V=3.10\pm0.10~{\rm GeV}^2.
\end{align}
\begin{table}[h]
\renewcommand{\arraystretch}{1.5}
\footnotesize
\begin{center}
\caption{The $B^+\to \omega$ TFFs $A_1(0), A_2(0)$ and $V(0)$ at large recoil point $q^2=0$. The predictions of other theories are used for comparison.}
\label{table:A1A2V0}
\begin{tabular}{l@{\hspace{2em}} l l l l}
\hline
  ~&$A_1(0)$~~~~~~~~~~ & $A_2(0)$~~~~~~~~~~ & $V(0)$\\
\hline
This work & $0.209^{+0.049}_{-0.042}$    & $0.206^{+0.051}_{-0.042}$    & $0.258^{+0.058}_{-0.048}$   \\
LCSR'04~\cite{Ball:2004rg}& $0.219$    & $0.198$    & $0.293$   \\
LCSR'15~\cite{Bharucha:2015bzk}     &  $0.243\pm0.031$    &  $0.270\pm0.040$    &  $0.304\pm0.038$ \\
HQEFT(I)~\cite{Wu:2006rd}     &  $0.221^{+0.012}_{-0.013}$    &  $0.211^{+0.011}_{-0.011}$    &  $0.275^{+0.014}_{-0.015}$ \\
HQEFT(II)~\cite{Wu:2006rd}     &  $0.214^{+0.013}_{-0.012}$    &  $0.170^{+0.010}_{-0.011}$    &  $0.268^{+0.014}_{-0.015}$ \\
SCET~\cite{Lu:2007sg}     &  $0.209$    &  $0.198$    &  $0.275$ \\
PQCD'02~\cite{Lu:2002ny}     &  $0.24\pm0.02$    &  $0.20\pm0.02$    &  $0.305\pm0.030$ \\
PQCD'09~\cite{Li:2009tx}     &  $0.15^{+0.03+0.02}_{-0.03-0.01}$    &  $0.12^{+0.03+0.02}_{-0.02-0.01}$    &  $0.19^{+0.04+0.03}_{-0.04-0.02}$ \\
LFQM~\cite{Verma:2011yw}     &  $0.23$    &  $0.21$    &  $0.27$ \\
CCQM~\cite{Soni:2020bvu}     &  $0.214\pm0.017$    &  $0.206\pm0.016$    &  $0.229\pm0.023$ \\
\hline
\end{tabular}
\end{center}
\end{table}
\begin{table}[h]
\renewcommand{\arraystretch}{1.5}
\footnotesize
\begin{center}
\caption{The uncertainty of TFFs at large recoil point caused by different input parameters.}
\label{table:TFFerror}
\begin{tabular}{l@{\hspace{2em}} l l l l}
\hline
  ~&$~A_1(0)$~~~~~~~~~~ & $~A_2(0)$~~~~~~~~~~ & $~V(0)$\\
\hline
$s_0$   &$^{+0.000}_{-0.000}$     &$^{+0.000}_{-0.000}$   &$^{+0.003}_{-0.000}$\\
$M^2$   &$^{+0.008}_{-0.007}$     &$^{+0.012}_{-0.010}$     &$^{+0.010}_{-0.003}$\\
$f_{B^+}$   &$^{+0.028}_{-0.022}$     &$^{+0.028}_{-0.021}$   &$^{+0.038}_{-0.024}$\\
$f_{\omega}$   &$^{+0.008}_{-0.008}$     &$^{+0.008}_{-0.008}$   &$^{+0.014}_{-0.007}$\\
$m_b$   &$^{+0.038}_{-0.033}$     &$^{+0.040}_{-0.033}$   &$^{+0.051}_{-0.037}$\\
$\phi _{2;\omega}^\|$   &$^{+0.006}_{-0.006}$     &$^{+0.007}_{-0.006}$   &$^{+0.009}_{-0.003}$\\
$\rm Total$   &$0.209^{+0.049}_{-0.042}$     &$0.206^{+0.051}_{-0.042}$   &$0.258^{+0.058}_{-0.048}$\\
\hline
\end{tabular}
\end{center}
\end{table}
Using above parameters and Eqs.~\eqref{Eq:A1TFFs}-\eqref{Eq:VTFFs}, we can first determine the $B^+\to \omega$ TFFs at a large recoil point, which are presented in Table~\ref{table:A1A2V0}. Meanwhile, it also includes the predictions from LCSR~\cite{Ball:2004rg,Bharucha:2015bzk}, HQEFT~\cite{Wu:2006rd}, SCET~\cite{Lu:2007sg}, PQCD~\cite{Lu:2002ny,Li:2009tx}, LFQM~\cite{Verma:2011yw}, and CCQM~\cite{Soni:2020bvu}. The HQEFT~\cite{Wu:2006rd} gives the results containing only the leading twist LCDA and the results promoted to twist-4 LCDA, respectively. We label the former as HQEFT(I) and the latter as HQEFT(II). As evident from this table, our predicted $A_1(0)$ shows agreement with HQEFT(II), SCET, and CCQM, while $A_2(0)$ is in good agreement with HQEFT(I), PQCD'02, LFQM, and CCQM. $V(0)$ has a certain gap with other theoretical groups. In order to understand the sources of TFFs uncertainty more clearly, we present the individual error contributions caused by different input parameters in Table~\ref{table:TFFerror}. Since the WW approximation is adopted for the twist-3 LCDAs, the error contribution from twist-2 LCDA $\phi _{2;\omega}^\|$ includes the contributions of the twist-3 LCDAs. Meanwhile, the nonlinear dependence of the three TFFs on the Borel parameter $M^2$ leads to asymmetric upper and lower bounds of uncertainty. Furthermore, there are two ratios $r_V=V(0)/A_1(0)$ and $r_2=A_2(0)/A_1(0)$ among the TFFs, which are the focuses of frequent experimental attention. These ratios are presented in Table~\ref{table:ratio} and include other theoretical predictions. Due to differences in methods and parameters used, there currently exists a certain disparity among different theoretical groups in predicting the large recoil point values of the three TFFs. We can further assess the reasonableness of the results by examining the overall trend of the TFFs.
\begin{figure}[h]
\begin{center}
\includegraphics[width=0.4\textwidth]{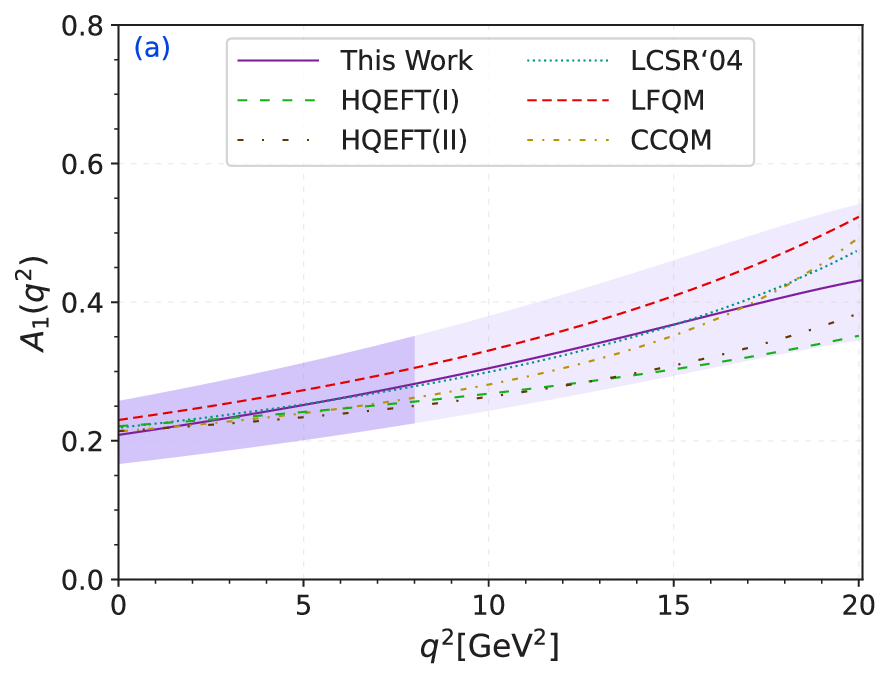}
\includegraphics[width=0.4\textwidth]{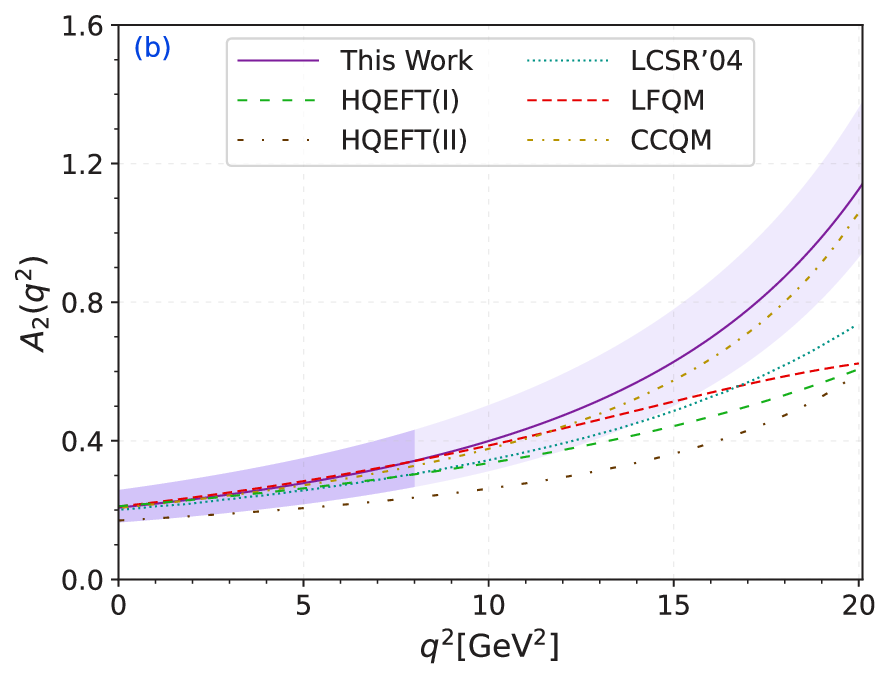}
\includegraphics[width=0.4\textwidth]{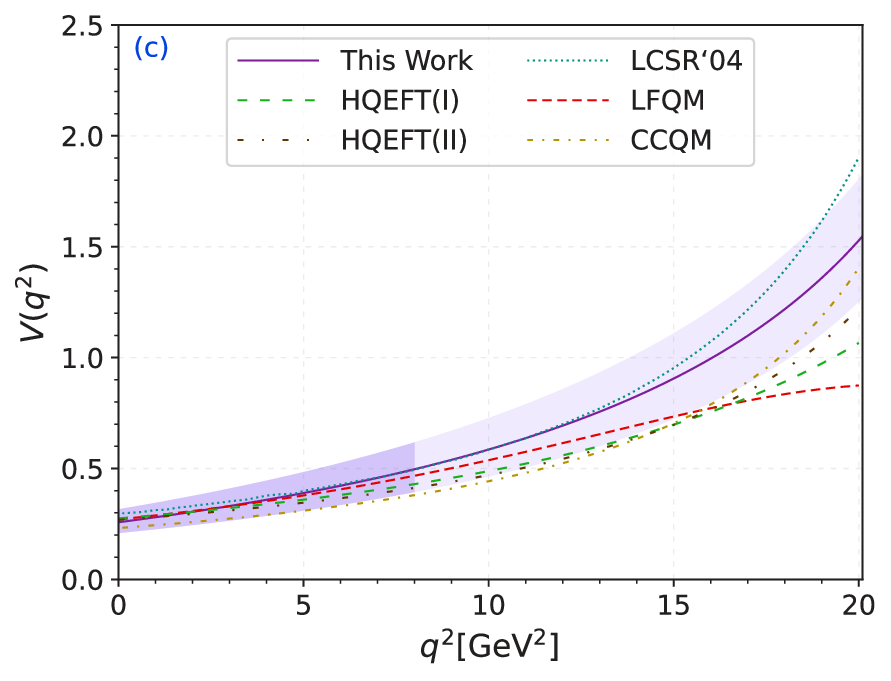}
\end{center}
\caption{The behavior of TFFs (a) $A_1(q^2)$, (b) $A_2(q^2)$, and (c) $V(q^2)$ in whole $q^2$ region. The solid line represents the central value and shaded bands corresponds to uncertainties. In addition, the darker shaded areas represent the results of LCSR method, while the lighter shaded areas represent the results of SSE method. For comparison, the predictions from LCSR'04~\cite{Ball:2004rg}, HQEFT~\cite{Wu:2006rd}, LFQM~\cite{Verma:2011yw} and CCQM~\cite{Soni:2020bvu} are also included.}
\label{Fig:TFFscomparison}
\end{figure}
\begin{table}[h]
\renewcommand{\arraystretch}{1.5}
\footnotesize
\begin{center}
\caption{The ratio $r_V$ and $r_2$ of $B^+\to \omega$ TFFs. For comparison, other theoretical results are also included.}
\label{table:ratio}
\begin{tabular}{l@{\hspace{2em}} l@{\hspace{2em}} l}
\hline
~  & $r_V$~~~~~~~~~ & $r_2$\\
\hline
This work& $1.234_{-0.322}^{+0.425}$    & $0.985_{-0.274}^{+0.347}$ \\
LCSR'04~\cite{Ball:2004rg}& $1.007$    & $0.904$     \\
LCSR'15~\cite{Bharucha:2015bzk}     &  $1.251^{+0.24}_{-0.21}$    &  $1.111^{+0.231}_{-0.207}$\\
HQEFT(I)~\cite{Wu:2006rd}     &  $1.244^{+0.100}_{-0.093}$    &  $0.955^{+0.077}_{-0.069}$ \\
HQEFT(II)~\cite{Wu:2006rd}    &  $1.252^{+0.099}_{-0.100}$    &  $0.794^{+0.066}_{-0.068}$ \\
SCET~\cite{Lu:2007sg}     &  $1.316$    &  $0.947$\\
PQCD'02~\cite{Lu:2002ny}     &  $1.271^{+0.170}_{-0.158}$    &  $0.833^{+0.112}_{-0.105}$  \\
PQCD'09~\cite{Li:2009tx}     &  $1.267^{+0.41+0.22}_{-0.34-0.20}$    &  $0.80^{+0.28+0.14}_{-0.19-0.11}$\\
LFQM~\cite{Verma:2011yw}     &  $1.17$    &  $0.91$ \\
CCQM~\cite{Soni:2020bvu}     &  $1.070^{+0.142}_{-0.133}$    &  $0.962\pm^{+0.112}_{-0.103}$\\
\hline
\end{tabular}
\end{center}
\end{table}

Since the decay width and branching fraction require the behavior of the TFFs across the entire physical $q^2$ region, and the LCSR method is applicable mainly to the low and intermediate $q^2$ region. To extrapolate the TFFs to all allowable physical region, $i.e.,$ $0\leq q^2\leq(m_{B^+}-m_\omega )^2$, the simplified series expansion (SSE) will be used~\cite{Bourrely:2008za}, $e.g.$,
\begin{align}
F_{i}(q^2)=\frac{1}{1-q^2/m^2_{R,i}}\sum_{k=0,1,2}\beta_{k,i} z^k(q^2,t_0),
\label{Eq:SSE}
\end{align}
where $F_{i}(q^2)$ with $i=(1,2,3)$ stands for TFFs $A_{1,2}(q^2)$ and $V(q^2)$, respectively. The function $z^k(q^2,t_0)=(\sqrt{t_+-q^2}-\sqrt{t_+-t_0})/(\sqrt{t_+-q^2}+\sqrt{t_+-t_0})$ with $t_{\pm}=(m_{B^+}\pm m_\omega )^2$ and $t_0=t_+(1-\sqrt{1-t_-/t_+})$. The masses of  low-lying $B^+$ meson resonance can be found in PDG~\cite{ParticleDataGroup:2024cfk}. In addition, the determination of real coefficients $\beta_{k,i}$ needs to satisfy the condition that the quality of extrapolation $\Delta$ is less than $1\%$, which is defined as:
\begin{align}
\Delta =\frac{\sum_t{|F_i(t)-F_i^{\rm fit}(t)|}}{\sum_t{|F_i(t)|}}~\times 100.
\label{eq:Delta}
\end{align}
The parameter of central value is exhibited in Table~\ref{table:fitParameter}.
\begin{table}[h]
\renewcommand{\arraystretch}{1.5}
\footnotesize
\begin{center}
\caption{The fitted parameters $m_{R,i}~(\rm{GeV})$, $\beta_{k,i}$ and quality of extrapolation $\Delta$ for $B^+\to \omega$ TFFs $A_1(0), A_2(0)$, and $V(0)$. All LCSR parameters are set to be their central values.}
\label{table:fitParameter}
\begin{tabular}{l@{\hspace{2em}} l l l l}
\hline
  ~&$~A_1(0)$~~~~~~~~~~ & $~A_2(0)$~~~~~~~~~~ & $~V(0)$\\
\hline
$m_{R,i}$   &$~5.726$     &~5.726   &~5.324\\
$\beta_{0,i}$   &~0.209     &~0.206   &~0.258\\
$\beta_{1,i}$   &-0.207     &-0.693  &-1.911\\
$\beta_{2,i}$   &-2.113    &~2.510   &-4.741\\
$\Delta$    &$~0.394\%$  &$~0.130\%$  &$~0.958\%$\\
\hline
\end{tabular}
\end{center}
\end{table}
Then, the specific behavior of $B^+\to \omega$ TFFs can be obtained, which is shown in Fig.~\ref{Fig:TFFscomparison}. Among them, the solid line represents our central result, and the shaded band indicates the uncertainty caused by the upper and lower limits of the input parameters. The results of the LCSR method are represented by darker shaded areas, and the SSE method are represented by lighter shaded areas. Meanwhile, the prediction from LCSR'04~\cite{Ball:2004rg}, HQEFT~\cite{Wu:2006rd}, LFQM~\cite{Verma:2011yw} and CCQM~\cite{Soni:2020bvu} are also presented. From Fig.~\ref{Fig:TFFscomparison}, we can observe that our predictions for the behavior of $A_1(q^2)$ align well with those of other groups within the uncertainties. $A_2(q^2)$ exhibits a trend that is in agreement with CCQM, while $V(q^2)$ demonstrates good consistency with LCSR'04. In the previous section, we mentioned that $A_1(q^2)$ contributes to all three helicity amplitudes $H_{\pm,0}(q^2)$, and this becomes even more evident in the high $q^2$ region. Therefore, in this case, we can expect that the next observables such as decay width, branching fraction, and CKM matrix element $|V_{ub}|$ may have good results, and these measurements can further test our theoretical predictions.

\begin{figure*}[t]
\begin{center}
\includegraphics[width=0.435\textwidth]{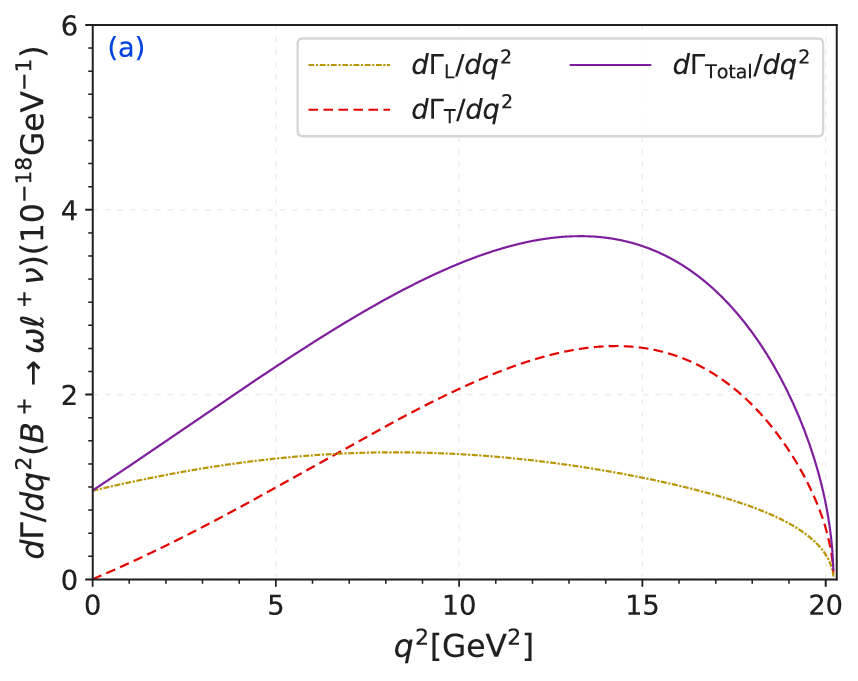}
\includegraphics[width=0.435\textwidth]{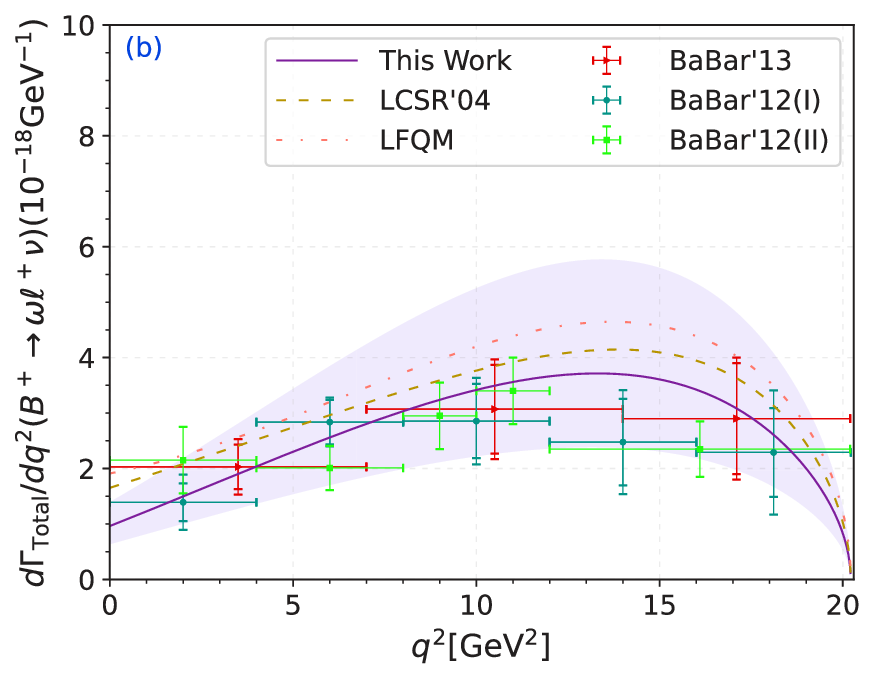}
\end{center}
\caption{The differential decay widths of $B^+\to \omega\ell^+\nu$ varying with the $q^2$: (a) central results for $d\Gamma_{\rm{L/T}}/dq^2$ and $d\Gamma_{\rm{Total}}/dq^2$; (b)
comparison of our $d\Gamma_{\rm{Total}}/dq^2$ with theoretical results of fitting and experimental data.}
\label{Fig:dGamma}
\end{figure*}

With the assistance of Eqs.~\eqref{Eq:dGammaL}-\eqref{Eq:dGammaT}, we can first obtain the trend of the differential decay widths as a function of $q^2$ by taking $|V_{ub}|=(3.67\pm 0.09\pm0.12)\times 10^{-3}$~\cite{HFLAV:2022esi}. The specific behavior is presented in Fig.~\ref{Fig:dGamma} ({\color{blue}a}). Naturally, the total differential decay width can also be determined and for comparison, the predicted results from \textit{BABAR}'12(I)~\cite{BaBar:2012thb}, \textit{BABAR}'12(II)~\cite{BaBar:2012dvs}, and \textit{BABAR}'13~\cite{BaBar:2013pls} are shown in Fig.~\ref{Fig:dGamma} ({\color{blue}b}). In view of the fact that in theory, few literatures directly give the data of differential decay width for $B^+\to \omega\ell^+\nu$, we decided to fit the TFFs data points of LCSR'04~\cite{Ball:2004rg}, LFQM~\cite{Verma:2011yw} with the help of Eq.~\eqref{Eq:SSE}. Finally, using the basic input parameters determined in this work, the prediction of differential decay width in these two literatures is obtained, which have already been presented in Fig.~\ref{Fig:dGamma} ({\color{blue}b}), in which it is clearly visible that our central results are in good agreement with the experimental data across different ranges of $q^2$. Compared to LCSR'04 and LFQM, the overall trend of the central results as a function of $q^2$ is generally consistent, and our results are closer to the experimental detections. After integrating over $q^2$ in entire physical region, the corresponding decay widths can be obtained,
\begin{align}
&\Gamma_{\rm{L}}(B^+\to \omega\ell^+\nu)=(2.29^{+3.76}_{-2.32})\times 10^{-17}~{\rm GeV},\nonumber\\
&\Gamma_{\rm{T}}(B^+\to \omega\ell^+\nu)=(3.16^{+1.32}_{-0.91})\times 10^{-17}~{\rm GeV},\nonumber\\
&\Gamma_{\rm{Total}}(B^+\to \omega\ell^+\nu)=(5.45^{+5.01}_{-2.78})\times 10^{-17}~{\rm GeV}.
\end{align}
Then, by using the Eq.~\eqref{Eq:Br} and lifetime $\tau_{B^+}=(1.638\pm0.004)$ ps, the branching fraction of $B^+\to \omega\ell^+\nu$ can also be calculated, which is listed in Table~\ref{table:Br}. In the uncertainty of our result, the first term represents the uncertainty of experimental measurement, while the second term corresponds to the uncertainty of theoretical calculation. The current experimental results primarily rely on the \textit{BABAR} and Belle Collaborations, with notable discrepancies observed between their most recent findings. Additionally, the fitting branching fraction results of LCSR'04~\cite{Ball:2004rg} and LFQM~\cite{Verma:2011yw} are also presented. Our results align well with those reported by \textit{BABAR}'13~\cite{BaBar:2013pls} and Belle'04~\cite{Belle:2004lza}. This table also shows that the current experimental measurements of the branching fraction of $B^+\to \omega\ell^+\nu$ decays have not been well unified, and we expect that these will be studied again by the experimental groups in the near future.

\begin{table}[h]
\renewcommand{\arraystretch}{1.5}
\footnotesize
\begin{center}
\caption{The prediction of the $B^+\to \omega\ell^+\nu$ branching fraction (in unit: $10^{-4}$). For comparison, the experimental and fitting theoretical results are also presented.}
\label{table:Br}
\begin{tabular}{l @{\hspace{60pt}}l}
\hline
  & $\mathcal{B}(B^+\to \omega\ell^+\nu)$ \\
\hline
This work    &  $1.35^{+0.02+1.22}_{-0.03-0.66}$    \\
\textit{BABAR}'13~\cite{BaBar:2013pls}     &  $1.35\pm0.21\pm0.11$    \\
\textit{BABAR}'12(I)~\cite{BaBar:2012thb}     &  $1.19\pm0.16\pm0.009$    \\
\textit{BABAR}'12(II)~\cite{BaBar:2012dvs}     &  $1.212\pm0.14\pm0.084$   \\
\textit{BABAR}'08~\cite{BaBar:2008vqc}     &  $1.14\pm0.16\pm0.08$    \\
Belle'13~\cite{Belle:2013hlo}     &  $1.07\pm0.16\pm0.07$    \\
Belle'04~\cite{Belle:2004lza}     &  $1.3\pm0.04\pm0.2\pm0.3$   \\
CLEO~\cite{CLEO:1992adu}     &  $<(1.6-2.7)$   \\
LCSR'04~\cite{Ball:2004rg}     &  $1.58$   \\
LFQM~\cite{Verma:2011yw}     &  $1.75$   \\
\hline
\end{tabular}
\end{center}
\end{table}
\begin{table}[h]
\renewcommand{\arraystretch}{1.5}
\footnotesize
\begin{center}
\caption{The prediction of $|V_{ub}|$ from $B^+\to \omega\ell^+\nu$ (in unit: $10^{-3}$). As a comparison, the different experimental results are listed.}
\label{table:Vub}
\begin{tabular}{l @{\hspace{60pt}}l}
\hline
  & $|V_{ub}|$ \\
\hline
This work    &  $3.66^{+0.12+1.26}_{-0.17-0.95}$    \\
PDG~\cite{ParticleDataGroup:2024cfk}          &  $3.82\pm0.20$\\
HFLAV~\cite{HFLAV:2022esi}      & $3.67\pm0.09\pm0.12$\\
FLAG~\cite{FlavourLatticeAveragingGroupFLAG:2024oxs}      &$3.63\pm0.16$\\
Belle'23~\cite{Belle:2023asx}   &$3.78\pm0.53$\\
Belle'13~\cite{Belle:2013hlo}   &$3.52\pm0.29$\\
Belle'10~\cite{Belle:2010hep}   &$3.43\pm0.33$\\
RBC/UKQCD'23~\cite{Flynn:2023nhi}  &$3.8\pm0.6$\\
RBC/UKQCD'15~\cite{Flynn:2015mha}  &$3.61\pm0.32$\\
CLEO~\cite{CLEO:2007vpk}        &$3.6^{+1.2}_{-1.0}$\\
JLQCD~\cite{Colquhoun:2022atw}    &$3.93\pm0.41$\\
Fermilab/MILC~\cite{FermilabLattice:2015mwy}  &$3.72\pm0.16$\\
UTfit~\cite{UTfit:2022hsi}    &$3.70\pm0.08$\\
\hline
\end{tabular}
\end{center}
\end{table}
Due to the significant discrepancies currently observed in the calculated values of $|V_{ub}|$ from different decay modes, we attempt to provide a theoretical prediction for $|V_{ub}|$ through the $B^+\rightarrow \omega \ell ^+\nu$ decay process. The following formula is required to be employed:
\begin{align}
|V_{ub}|=\sqrt{\frac{\mathcal{B}_{\rm Exp}(B^+\rightarrow \omega \ell ^+\nu)}{\tau_{B^+}A}},
\label{Eq:Vub}
\end{align}
with A represents the differential decay width independent of $|V_{ub}|$,
\begin{align}
A=\frac{1}{|V_{ub}|^2}\int_0^{q^2_{\rm{max}}}\frac{d\Gamma (B^+\rightarrow \omega \ell ^+\nu)}{dq^2}.
\end{align}
Here, we adopt the most recent experimental result of $\mathcal{B}_{\rm Exp}(B^+\rightarrow \omega \ell ^+\nu)=(1.35\pm0.21\pm0.11)\times 10^{-4}$ from \textit{BABAR}'13~\cite{BaBar:2013pls}. Then, we can get
\begin{align}
|V_{ub}|=(3.66^{+0.12+1.26}_{-0.17-0.95})\times 10^{-3}.
\end{align}
For the purpose of clear result comparison, we have compiled the results from PDG~\cite{ParticleDataGroup:2024cfk}, HFLAV~\cite{HFLAV:2022esi}, FLAG~\cite{FlavourLatticeAveragingGroupFLAG:2024oxs}, Belle'23~\cite{Belle:2023asx}, Belle'13~\cite{Belle:2013hlo}, Belle'10~\cite{Belle:2010hep}, RBC/UKQCD'23~\cite{Flynn:2023nhi}, RBC/UKQCD'15~\cite{Flynn:2015mha}, CLEO~\cite{CLEO:2007vpk}, JLQCD~\cite{Colquhoun:2022atw}, Fermilab/MILC~\cite{FermilabLattice:2015mwy}, and UDfit~\cite{UTfit:2022hsi} in Table~\ref{table:Vub}. Our result has agreement with HFLAV, FLAG, RBC/UKQCD'15, CLEO and UTfit. Among them, these results are mainly determined by the semileptonic decay $B\to \pi\ell\nu$. And we predict similar results from the semileptonic decay of $B^+\to \omega\ell^+\nu$, which also shows that our current theoretical calculation are reasonable and have the potential to provide a reference for future experiments in this decay channel.

\section{Summary}\label{Sec:4}
In the framework of the LCSR method, we have calculated the semileptonic decay $B^+\to \omega\ell^+\nu$. Firstly, for twist-2 LCDA $\phi _{2;\omega}^\| (x,\mu)$, we construct the LCHO model to describe its behavior, whose model parameters are determined by additional three conditions. The Fig.~\ref{Fig:DA} shows that LCHO model of $\phi _{2;\omega}^\| (x,\mu_0)$ has double-peak behavior. And the Fig.~\ref{Fig:twist3DA} also shows the twist-3 LCDAs $\phi _{3;\omega}^{\bot}(x,\mu_0)$ and $\psi _{3;\omega}^{\bot}(x,\mu_0)$ under the WW approximation. Our predictions are very close to QCDSR(WW)~\cite{Ball:1998sk}. The results of $A_1(0)$, $A_2(0)$, and $V(0)$ at large recoil point are listed in Table~\ref{table:A1A2V0}. Meanwhile, for the convenience of the experimental group, two important ratios $r_V$ and $r_2$ are also presented in Table ~\ref{table:ratio}. Due to the different methods and the lack of experimental evidence, the predictions of different theoretical groups for these two ratios cannot reach a high degree of consistency.

After extrapolating the TFFs to whole physical $q^2$ region, the complete behavior of TFFs are shown in Fig.~\ref{Fig:TFFscomparison}. Our predictions for the overall trends of $A_1(q^2)$, $A_2(q^2)$, and $V(q^2)$ each show good consistency with those of other theoretical groups, particularly for $A_1(q^2)$. Immediately following that, some interesting physical observables of $B^+\to \omega\ell^+\nu$ can be determined. Firstly, the differential decay width as a function of $q^2$ has been shown in Fig.~\ref{Fig:dGamma}, which shows excellent agreement with those of \textit{BABAR}~\cite{BaBar:2012dvs,BaBar:2013pls,BaBar:2012thb} within different $q^2$ region. Secondly, by integrating over $q^2$ from 0 to $(m_{B^+}-m_\omega )^2$, the branching fraction is given in Table~\ref{table:Br}, which indicates our prediction aligns well with \textit{BABAR}'13~\cite{BaBar:2013pls} and Belle'04~\cite{Belle:2004lza}. Finally, with the help of $\mathcal{B}_{\rm Exp}(B^+\to \omega\ell^+\nu)=(1.35\pm0.21\pm0.11)\times 10^{-4}$ from \textit{BABAR}'13~\cite{BaBar:2013pls}, we extract the CKM matrix element $|V_{ub}|=(3.66^{+0.12+1.26}_{-0.17-0.95})\times 10^{-3}$ and compare it with other experimental results in Table~\ref{table:Vub}. Our predictions are consistent with the average results from HFLAV and FLAG, as well as with measurements from other experimental collaborations.

Overall, our calculated results show good consistency with others. Currently, the experimental data for $B^+\to \omega\ell^+\nu$ detection are still not abundant. As part of the semileptonic decay process of $B$-mesons, it is believed that this process will be investigated again in the near future.

\section{Acknowledgments}

Hai-Bing Fu and Tao Zhong would like to thank the Institute of High Energy Physics of Chinese Academy of Sciences for their warm and kind hospitality. This work was supported in part by the National Natural Science Foundation of China under Grant No.12265010, No.12265009, the Project of Guizhou Provincial Department of Science and Technology under Grants No.MS[2025]219, No.CXTD[2025]030, and No.ZK[2023]024.


\begin{thebibliography}{99}

\bibitem{Seng:2018yzq}
C.~Y.~Seng, M.~Gorchtein, H.~H.~Patel, and M.~J.~Ramsey-Musolf,
Reduced hadronic uncertainty in the determination of $|V_{ud}|$,
\href{https://doi.org/10.1103/PhysRevLett.121.241804}
{Phys. Rev. Lett. \textbf{121}, 241804 (2018)}.

\bibitem{ParticleDataGroup:2024cfk}
S.~Navas \textit{et al.} (Particle Data Group),
Review of particle physics,
\href{https://doi.org/10.1103/PhysRevD.110.030001}
{Phys. Rev. D \textbf{110}, 030001 (2024)}.

\bibitem{BaBar:2010efp}
P.~del Amo Sanchez \textit{et al.} (\textit{BABAR} Collaboration),
Study of $B \to \pi \ell \nu$ and $B \to \rho \ell \nu$ decays and determination of $|V_{ub}|$,
\href{https://doi.org/10.1103/PhysRevD.83.032007}
{Phys. Rev. D \textbf{83}, 032007 (2011)}.
[\href{https://arxiv.org/abs/1005.3288}
{arXiv:1005.3288}]

\bibitem{Belle:2006hlt}
T.~Hokuue \textit{et al.} (Belle Collaboration),
Measurements of branching fractions and $q^2$ distributions for $B\to \pi \ell \nu$ and $B\to \rho \ell \nu$ decays with $B\to D^{(*)} \ell\nu$ decay tagging,
\href{https://doi.org/10.1016/j.physletb.2007.02.067}
{Phys. Lett. B \textbf{648}, 139 (2007)}.

\bibitem{CLEO:1999mif}
B.~H.~Behrens \textit{et al.} (CLEO Collaboration),
Measurement of $B\to \rho\ell\nu$ decay and $|V_{ub}|$,
\href{https://doi.org/10.1103/PhysRevD.61.052001}
{Phys. Rev. D \textbf{61}, 052001 (2000)}.

\bibitem{BaBar:2010npl}
P.~del Amo Sanchez \textit{et al.} (\textit{BABAR} Collaboration),
Measurement of the $B^0 \to \pi^- \ell^+ \nu$ and $B^+ \to \eta^{(')} \ell^+ \nu$ branching fractions, the $B^0 \to \pi^- \ell^+ \nu$ and $B^+ \to \eta \ell^+ \nu$ form-factor shapes, and determination of $|V_{ub}|$,
\href{https://doi.org/10.1103/PhysRevD.83.052011}
{Phys. Rev. D \textbf{83}, 052011 (2011)}.

\bibitem{LHCb:2020ist}
R.~Aaij \textit{et al.} (LHCb Collaboration),
First observation of the decay $B_s^0 \to K^-\mu^+\nu_\mu$ and measurement of $|V_{ub}|/|V_{cb}|$,
\href{https://doi.org/10.1103/PhysRevLett.126.081804}
{Phys. Rev. Lett. \textbf{126}, 081804 (2021)}.
[\href{https://arxiv.org/abs/2012.05143}
{arXiv:2012.05143}]




\bibitem{ARGUS:1990amj}
H.~Albrecht \textit{et al.} (ARGUS Collaboration),
Reconstruction of semileptonic $b\to u$ decays,
\href{https://doi.org/10.1016/0370-2693(91)90251-K}
{Phys. Lett. B \textbf{255}, 297 (1991)}.

\bibitem{CLEO:1992adu}
A.~Bean \textit{et al.} (CLEO Collaboration),
A search for exclusive $b\to u$ semileptonic decays of $B$ mesons,
\href{https://doi.org/10.1103/PhysRevLett.70.2681}
{Phys. Rev. Lett. \textbf{70}, 2681 (1993)}.


\bibitem{Belle:2004lza}
C.~Schwanda \textit{et al.} (Belle Collaboration),
Evidence for $B^+\to \omega \ell^+ \nu,$
\href{https://doi.org/10.1103/PhysRevLett.93.131803}
{Phys. Rev. Lett. \textbf{93}, 131803 (2004)}.



\bibitem{BaBar:2013pls}
J.~P.~Lees \textit{et al.} (\textit{BABAR} Collaboration),
Measurement of the $B^+ \to \omega \ell^+ \nu$ branching fraction with semileptonically tagged $B$ mesons,
\href{https://doi.org/10.1103/PhysRevD.88.072006}
{Phys. Rev. D \textbf{88}, 072006 (2013)}.

\bibitem{BaBar:2008vqc}
B.~Aubert \textit{et al.} (\textit{BABAR} Collaboration),
Measurement of the $B^{+} \to \omega \ell^{+} \nu$ and $B^{+}\to \eta \ell^{+} \nu$ branching fractions,
\href{https://doi.org/10.1103/PhysRevD.79.052011}
{Phys. Rev. D \textbf{79}, 052011 (2009)}.


\bibitem{BaBar:2012thb}
J.~P.~Lees \textit{et al.} (\textit{BABAR} Collaboration),
Branching fraction and form-factor shape measurements of exclusive charmless semileptonic $B$ decays, and determination of $|V_{ub}|$,
\href{https://doi.org/10.1103/PhysRevD.86.092004}
{Phys. Rev. D \textbf{86}, 092004 (2012)}.

\bibitem{BaBar:2012dvs}
J.~P.~Lees \textit{et al.} (\textit{BABAR} Collaboration),
Branching fraction measurement of $B^+ \to \omega \ell^+ \nu$ decays,
\href{https://doi.org/10.1103/PhysRevD.87.032004}
{Phys. Rev. D \textbf{87}, 032004 (2013)}.


\bibitem{Belle:2013hlo}
A.~Sibidanov \textit{et al.} (Belle Collaboration),
Study of exclusive $B \to X_u \ell \nu$ decays and extraction of $|V_{ub}|$ using full reconstruction tagging at the Belle experiment,
\href{https://doi.org/10.1103/PhysRevD.88.032005}
{Phys. Rev. D \textbf{88}, 032005 (2013)}.

\bibitem{Bharucha:2015bzk}
A.~Bharucha, D.~M.~Straub, and R.~Zwicky,
$B\to V\ell^+\ell^-$ in the standard model from light-cone sum rules,
\href{https://doi.org/10.1007/JHEP08(2016)098}
{J. High Energy Phys. \textbf{08}, (2016) 098}.

\bibitem{Ball:2004rg}
P.~Ball and R.~Zwicky,
$B_{d,s} \to  \rho, \omega, K^*, \phi$ decay form-factors from light-cone sum rules revisited,
\href{https://doi.org/10.1103/PhysRevD.71.014029}
{Phys. Rev. D \textbf{71}, 014029 (2005)}.


\bibitem{Wu:2006rd}
Y.~L.~Wu, M.~Zhong, and Y.~B.~Zuo,
$B_{(s)}, D_{(s)}\to \pi, K, \eta, \rho, K^*, \omega, \phi$ transition form factors and decay rates with extraction of the CKM parameters $|V_{ub}|, |V_{cs}|, |V_{cd}|$,
\href{https://doi.org/10.1142/S0217751X06033209}
{Int. J. Mod. Phys. A \textbf{21}, 6125 (2006)}.


\bibitem{Lu:2007sg}
C.~D.~Lu, W.~Wang, and Z.~T.~Wei,
Heavy-to-light form factors on the light-cone,
\href{ https://doi.org/10.1103/PhysRevD.76.014013}
{Phys. Rev. D \textbf{76}, 014013 (2007)}.


\bibitem{Lu:2002ny}
C.~D.~Lu and M.~Z.~Yang,
$B$ to light meson transition form-factors calculated in perturbative QCD approach,
\href{https://doi.org/10.1140/epjc/s2003-01199-y}
{Eur. Phys. J. C \textbf{28}, 515 (2003)}.

\bibitem{Li:2009tx}
R.~H.~Li, C.~D.~Lu, and W.~Wang,
Transition form factors of $B$ decays into $p$-wave axial-vector mesons in the perturbative QCD approach,
\href{https://doi.org/10.1103/PhysRevD.79.034014}
{Phys. Rev. D \textbf{79}, 034014 (2009)}.

\bibitem{Verma:2011yw}
R.~C.~Verma,
Decay constants and form factors of $s$-wave and $p$-wave mesons in the covariant light-front quark model,
\href{https://iopscience.iop.org/article/10.1088/0954-3899/39/2/025005}
{J. Phys. G \textbf{39}, 025005 (2012)}.

\bibitem{Soni:2020bvu}
N.~R.~Soni, A.~Issadykov, A.~N.~Gadaria, J.~J.~Patel, and J.~N.~Pandya,
Rare $b \rightarrow d$ decays in covariant confined quark model,
\href{https://doi.org/10.1140/epja/s10050-022-00685-y}
{Eur. Phys. J. A \textbf{58}, 39 (2022)}.

\bibitem{Balitsky:1989ry}
I.~I.~Balitsky, V.~M.~Braun, and A.~V.~Kolesnichenko,
Radiative decay $\sigma^+\to  \rho \gamma$ in quantum chromodynamics,
\href{https://doi.org/10.1016/0550-3213(89)90570-1}
{Nucl. Phys. B\textbf{312}, 509 (1989)}.

\bibitem{Chernyak:1990ag}
V.~L.~Chernyak and I.~R.~Zhitnitsky,
$B$ meson exclusive decays into baryons,
\href{https://doi.org/10.1016/0550-3213(90)90612-H}
{Nucl. Phys. B\textbf{345}, 137 (1990)}.

\bibitem{Ball:2004ye}
P.~Ball and R.~Zwicky,
New results on $B \to \pi, K, \eta$ decay formfactors from light-cone sum rules,
\href{https://doi.org/10.1103/PhysRevD.71.014015}
{Phys. Rev. D \textbf{71}, 014015 (2005)}.


\bibitem{Ball:2001fp}
P.~Ball and R.~Zwicky,
Improved analysis of $B\to \pi e \nu$ from QCD sum rules on the light-cone,
\href{https://iopscience.iop.org/article/10.1088/1126-6708/2001/10/019}
{J. High Energy Phys. \textbf{10}, (2001) 019}.

\bibitem{Fu:2018yin}
H.~B.~Fu, L.~Zeng, R.~L\"u, W.~Cheng, and X.~G.~Wu,
The $D\to \rho$ semileptonic and radiative decays within the light-cone sum rules,
\href{https://doi.org/10.1140/epjc/s10052-020-7758-4}
{Eur. Phys. J. C \textbf{80}, 194 (2020)}.


\bibitem{Cheng:2017bzz}
W.~Cheng, X.~G.~Wu, and H.~B.~Fu,
Reconsideration of the $B \to K^*$ transition form factors within the QCD light-cone sum rules,
\href{https://doi.org/10.1103/PhysRevD.95.094023}
{Phys. Rev. D \textbf{95}, 094023 (2017)}.


\bibitem{Aliev:2004vf}
T.~M.~Aliev, M.~Savci, and A.~Ozpineci,
Form factors of $D_s^+\to \phi\bar{\ell}\nu$ decay in QCD light-cone sum rule,
\href{https://doi.org/10.1140/epjc/s2004-02034-9}
{Eur. Phys. J. C \textbf{38}, 85 (2004)}.



\bibitem{Hu:2024tmc}
D.~D.~Hu, X.~G.~Wu, L.~Zeng, H.~B.~Fu, and T.~Zhong,
Improved light-cone harmonic oscillator model for the $\phi$-meson longitudinal leading-twist light-cone distribution amplitude and its effects to $D_s^+\to \phi\ell^+\nu_{\ell}$,
\href{https://doi.org/10.1103/PhysRevD.110.056017}
{Phys. Rev. D \textbf{110}, 056017 (2024)}.


\bibitem{BESIII:2015kin}
M.~Ablikim \textit{et al.} (BESIII Collaboration),
Measurement of the form factors in the decay $D^+ \to \omega e^+ \nu_{e}$ and search for the decay $D^+ \to \phi e^+ \nu_{e}$,
\href{https://doi.org/10.1103/PhysRevD.92.071101}
{Phys. Rev. D \textbf{92}, 071101 (2015)}.


\bibitem{Gronau:2009mp}
M.~Gronau and J.~L.~Rosner,
$\omega - \phi$ mixing and weak annihilation in $D_s$ decays,
\href{https://doi.org/10.1103/PhysRevD.79.074006}
{Phys. Rev. D \textbf{79}, 074006 (2009)}.


\bibitem{Li:2020ylu}
H.~B.~Li and M.~Z.~Yang,
Semileptonic decay of $D_s^+\to \pi^0 \ell^+ \nu_\ell$ via neutral meson mixing,
\href{https://doi.org/10.1016/j.physletb.2020.135879}
{Phys. Lett. B \textbf{811}, 135879 (2020)}.


\bibitem{Kucukarslan:2006wk}
A.~Kucukarslan and U.~G.~Meissner,
$\omega-\phi$ mixing in chiral perturbation theory,
\href{https://doi.org/10.1142/S0217732306020743}
{Mod. Phys. Lett. A \textbf{21}, 1423 (2006)}.



\bibitem{Ambrosino:2009sc}
F.~Ambrosino, A.~Antonelli, and S.~Bocchetta, \textit{et al.} (KLOE Collaboration),
A Global fit to determine the pseudoscalar mixing angle and the gluonium content of the $\eta^{\prime}$ meson,
\href{https://iopscience.iop.org/article/10.1088/1126-6708/2009/07/105}
{J. High Energy Phys. \textbf{07}, 105 (2009)}.

\bibitem{Klingl:1996by}
F.~Klingl, N.~Kaiser, and W.~Weise,
Effective Lagrangian approach to vector mesons, their structure and decays,
\href{https://doi.org/10.1007/s002180050167}
{Z. Phys. A \textbf{356}, 193 (1996)}.


\bibitem{Benayoun:2007cu}
M.~Benayoun, P.~David, L.~DelBuono, O.~Leitner, and H.~B.~O'Connell,
The dipion mass spectrum In $e^+ e^-$ annihilation and tau decay: A dynamical $(\rho, \omega, \phi)$ mixing approach,
\href{https://doi.org/10.1140/epjc/s10052-008-0586-6}
{Eur. Phys. J. C \textbf{55}, 199 (2008)}.


\bibitem{Benayoun:2009im}
M.~Benayoun, P.~David, L.~DelBuono, and O.~Leitner,
A global treatment Of VMD physics up to the $\phi$: I. $e^+ e^-$ annihilations, anomalies and vector meson partial widths,
\href{https://doi.org/10.1140/epjc/s10052-009-1197-6}
{Eur. Phys. J. C \textbf{65}, 211 (2010)}.


\bibitem{Choi:2015ywa}
H.~M.~Choi, C.~R.~Ji, Z.~Li, and H.~Y.~Ryu,
Variational analysis of mass spectra and decay constants for ground state pseudoscalar and vector mesons in the light-front quark model,
\href{https://doi.org/10.1103/PhysRevC.92.055203}
{Phys. Rev. C \textbf{92}, 055203 (2015)}.


\bibitem{Ball:1998sk}
P.~Ball, V.~M.~Braun, Y.~Koike, and K.~Tanaka,
Higher twist distribution amplitudes of vector mesons in QCD: Formalism and twist three distributions,
\href{https://doi.org/10.1016/S0550-3213(98)00356-3}
{Nucl. Phys. B\textbf{529}, 323 (1998)}.
P.~Ball and V.~M.~Braun,
Higher twist distribution amplitudes of vector mesons in QCD: Twist-4 distributions and meson mass corrections,
\href{https://doi.org/10.1016/S0550-3213(99)00014-0}
{Nucl. Phys. B\textbf{543}, 201 (1999)}.

\bibitem{Huang:1998gp}
T.~Huang and Z.~H.~Li,
$B\to K^*\gamma$ in the light-cone QCD sum rule,
\href{https://doi.org/10.1103/PhysRevD.57.1993}
{Phys. Rev. D \textbf{57}, 1993 (1998)}.

\bibitem{Huang:2001xb}
T.~Huang, Z.~H.~Li, and X.~Y.~Wu,
Improved approach to the heavy to light form-factors in the light-cone QCD sum rules,
\href{https://doi.org/10.1103/PhysRevD.63.094001}
{Phys. Rev. D \textbf{63}, 094001 (2001)}.

\bibitem{Wang:2002aje}
Z.~G.~Wang, M.~Z.~Zhou, and T.~Huang,
$B\to \pi$ weak form-factor with chiral current in the light-cone sum rules,
\href{https://doi.org/10.1103/PhysRevD.67.094006}
{Phys. Rev. D \textbf{67}, 094006 (2003)}.

\bibitem{Fu:2014pba}
H.~B.~Fu, X.~G.~Wu, H.~Y.~Han, and Y.~Ma,
$B\to \rho$ transition form factors and the $\rho$-meson transverse leading-twist distribution amplitude,
\href{https://iopscience.iop.org/article/10.1088/0954-3899/42/5/055002}
{J. Phys. G \textbf{42}, 055002 (2015)}.


\bibitem{LatticeParton:2022zqc}
J.~Hua \textit{et al.} (Lattice Parton Collaboration),
Pion and kaon distribution amplitudes from lattice QCD,
\href{ https://doi.org/10.1103/PhysRevLett.129.132001}
{Phys. Rev. Lett. \textbf{129} (2022), 132001 (2022)}.

\bibitem{Hua:2020gnw}
J.~Hua \textit{et al.} (Lattice Parton Collaboration),
Distribution amplitudes of $K^*$ and $\phi$ at the physical pion mass from lattice QCD,
\href{ https://doi.org/10.1103/PhysRevLett.127.062002}
{Phys. Rev. Lett. \textbf{127}, 062002  (2021)}.





\bibitem{Zhong:2011rg}
T.~Zhong, X.~G.~Wu, H.~Y.~Han, Q.~L.~Liao, H.~B.~Fu, and Z.~Y.~Fang,
Revisiting the twist-3 distribution amplitudes of $K$ meson within the QCD background field approach,
\href{https://iopscience.iop.org/article/10.1088/0253-6102/58/2/16}
{Commun. Theor. Phys. \textbf{58}, 261 (2012)}.


\bibitem{Wu:2007rt}
X.~G.~Wu, T.~Huang, and Z.~Y.~Fang,
$B\to K$ transition form-factor up to $\mathcal{O}(1/m_b^2)$ within the $k_T$ factorization approach,
\href{https://doi.org/10.1140/epjc/s10052-007-0421-5}
{Eur. Phys. J. C \textbf{52}, 561 (2007)}.


\bibitem{Huang:2004su}
T.~Huang and X.~G.~Wu,
A model for the twist-3 wave function of the pion and its contribution to the pion form-factor,
\href{https://doi.org/10.1103/PhysRevD.70.093013}
{Phys. Rev. D \textbf{70}, 093013 (2004)}.


\bibitem{Zhong:2022ecl}
T.~Zhong, H.~B.~Fu, and X.~G.~Wu,
Investigating the ratio of CKM matrix elements $|V_{ub}|/|V_{cb}|$ from semileptonic decay $B_s^0\to K^-\mu^+\nu_{\mu}$ and kaon twist-2 distribution amplitude,
\href{https://doi.org/10.1103/PhysRevD.105.116020}
{Phys. Rev. D \textbf{105}, 116020 (2022)}.



\bibitem{Zhong:2021epq}
T.~Zhong, Z.~H.~Zhu, H.~B.~Fu, X.~G.~Wu, and T.~Huang,
Improved light-cone harmonic oscillator model for the pionic leading-twist distribution amplitude,
\href{https://doi.org/10.1103/PhysRevD.104.016021}
{Phys. Rev. D \textbf{104}, 016021 (2021)}.

\bibitem{Hu:2023pdl}
D.~D.~Hu, X.~G.~Wu, H.~B.~Fu, T.~Zhong, Z.~H.~Wu, and L.~Zeng,
Properties of the $\eta _q$ leading-twist distribution amplitude and its effects to the $B/D^+ \to \eta ^{(\prime )}\ell ^+ \nu _\ell $ decays,
\href{https://doi.org/10.1140/epjc/s10052-023-12333-w}
{Eur. Phys. J. C \textbf{84}, 15 (2024)}.


\bibitem{Huang:2022xny}
D.~Huang, T.~Zhong, H.~B.~Fu, Z.~H.~Wu, X.~G.~Wu, and H.~Tong,
$K_0^*(1430)$ twist-2 distribution amplitude and $B_s,D_s \to K_0^*(1430)$ transition form factors,
\href{https://doi.org/10.1140/epjc/s10052-023-11851-x}
{Eur. Phys. J. C \textbf{83}, 680 (2023)}.



\bibitem{Yang:2024ang}
Y.~L.~Yang, H.~J.~Tian, Y.~X.~Wang, H.~B.~Fu, T.~Zhong, S.~Q.~Wang, and D.~Huang,
Probing $|V_{cs}|$ and lepton flavor universality through $D\to K_0^*(1430)\ell\nu_{\ell}$ decays,
\href{https://doi.org/10.1103/PhysRevD.110.116030}
{Phys. Rev. D \textbf{110}, 116030 (2024)}.


\bibitem{Wu:2022qqx}
Z.~H.~Wu, H.~B.~Fu, T.~Zhong, D.~Huang, D.~D.~Hu, and X.~G.~Wu,
$a_0(980)$-meson twist-2 distribution amplitude within the QCD sum rules and investigation of $D\to a_0(980)(\to \eta\pi)e^+\nu_e$,
\href{https://doi.org/10.1016/j.nuclphysa.2023.122671}
{Nucl. Phys. A\textbf{1036}, 122671 (2023)}.


\bibitem{Fu:2016yzx}
H.~B.~Fu, X.~G.~Wu, W.~Cheng, and T.~Zhong,
$\rho$-meson longitudinal leading-twist distribution amplitude within QCD background field theory,
\href{https://doi.org/10.1103/PhysRevD.94.074004}
{Phys. Rev. D \textbf{94}, 074004 (2016)}.



\bibitem{Fu:2014uea}
H.~B.~Fu, X.~G.~Wu, and Y.~Ma,
$B\to K^*$ Transition form factors and the semi-leptonic decay $B \to K^* \mu^+ \mu^-$,
\href{https://iopscience.iop.org/article/10.1088/0954-3899/43/1/015002}
{J. Phys. G \textbf{43}, 015002 (2016)}.


\bibitem{Gao:2019lta}
J.~Gao, C.~D.~L\"u, Y.~L.~Shen, Y.~M.~Wang, and Y.~B.~Wei,
Precision calculations of $B \to V$ form factors from soft-collinear effective theory sum rules on the light-cone,
\href{https://doi.org/10.1103/PhysRevD.101.074035}
{Phys. Rev. D \textbf{101}, 074035 (2020)}.

\bibitem{Gilani:2003hf}
A.~H.~S.~Gilani, Riazuddin and T.~A.~Al-Aithan,
Ward identities, $B\to \rho$ form-factors and $|V_{ub}|$,
\href{https://iopscience.iop.org/article/10.1088/1126-6708/2003/09/065}
{J. High Energy Phys. \textbf{09}, (2003) 065}.


\bibitem{Fu:2014cna}
H.~B.~Fu, X.~G.~Wu, H.~Y.~Han, Y.~Ma, and H.~Y.~Bi,
The $\rho$-meson longitudinal leading-twist distribution amplitude,
\href{https://doi.org/10.1016/j.physletb.2014.09.055}
{Phys. Lett. B \textbf{738}, 228 (2014)}.


\bibitem{Kaur:2020emh}
S.~Kaur, C.~Mondal and H.~Dahiya,
Light-front holographic $\rho$-meson distributions in the momentum space,
\href{https://doi.org/10.1007/JHEP01(2021)136}
{J. High Energy Phys. \textbf{01} (2021) 136}.


\bibitem{Guo:1991eb}
X.~H.~Guo and T.~Huang,
Hadronic wave functions in $D$ and $B$ decays,
\href{ https://doi.org/10.1103/PhysRevD.43.2931}
{Phys. Rev. D \textbf{43}, 2931 (1991)}.

\bibitem{Dimou:2012un}
M.~Dimou, J.~Lyon, and R.~Zwicky,
Exclusive chromomagnetism in heavy-to-light FCNCs,
\href{https://doi.org/10.1103/PhysRevD.87.074008}
{Phys. Rev. D \textbf{87}, 074008 (2013)}.


\bibitem{Arthur:2010xf}
R.~Arthur, P.~A.~Boyle, D.~Brommel, M.~A.~Donnellan, J.~M.~Flynn, A.~Juttner, T.~D.~Rae, and C.~T.~C.~Sachrajda,
Lattice results for low moments of light meson distribution amplitudes,
\href{https://doi.org/10.1103/PhysRevD.83.074505}
{Phys. Rev. D \textbf{83}, 074505 (2011)}.


\bibitem{Ball:2007rt}
P.~Ball and G.~W.~Jones,
Twist-3 distribution amplitudes of $K^*$ and $\phi$ mesons,
\href{https://iopscience.iop.org/article/10.1088/1126-6708/2007/03/069}
{J. High Energy Phys. \textbf{03}, (2007) 069}.


\bibitem{Braun:2003rp}
V.~M.~Braun, G.~P.~Korchemsky, and D.~M\"uller,
The uses of conformal symmetry in QCD,
\href{https://doi.org/10.1016/S0146-6410(03)90004-4}
{Prog. Part. Nucl. Phys. \textbf{51}, 311 (2003)}.

\bibitem{Bourrely:2008za}
C.~Bourrely, I.~Caprini, and L.~Lellouch,
Model-independent description of $B\to \pi \ell \nu$ decays and a determination of $|V_{ub}|$,
\href{https://doi.org/10.1103/PhysRevD.82.099902}
{Phys. Rev. D \textbf{79}, 013008 (2009)}.






\bibitem{HFLAV:2022esi}
Y.~S.~Amhis \textit{et al.} (HFLAV Collaboration),
Averages of $b$-hadron, $c$-hadron, and $\tau$-lepton properties as of 2021,
\href{https://doi.org/10.1103/PhysRevD.107.052008}
{Phys. Rev. D \textbf{107}, 052008 (2023)}.


\bibitem{FlavourLatticeAveragingGroupFLAG:2024oxs}
Y.~Aoki \textit{et al.} (Flavour Lattice Averaging Group (FLAG)), FLAG Review 2024,
[\href{https://arxiv.org/abs/2411.04268}
{arXiv:2411.04268}].


\bibitem{Belle:2023asx}
L.~Cao \textit{et al.} (Belle Collaboration),
First simultaneous determination of inclusive and exclusive $|V_{ub}|$,
\href{https://doi.org/10.1103/PhysRevLett.131.211801}
{Phys. Rev. Lett. \textbf{131}, 211801 (2023)}.

\bibitem{Belle:2010hep}
H.~Ha \textit{et al.} (Belle Collaboration),
Measurement of the decay $B^0\to\pi^-\ell^+\nu$ and determination of $|V_{ub}|$,
\href{https://doi.org/10.1103/PhysRevD.83.071101}
{Phys. Rev. D \textbf{83}, 071101 (2011)}.

\bibitem{Flynn:2023nhi}
J.~M.~Flynn \textit{et al.} (RBC/UKQCD Collaboration),
Exclusive semileptonic $B_s\to K \ell\nu$ decays on the lattice,
\href{https://doi.org/10.1103/PhysRevD.107.114512}
{Phys. Rev. D \textbf{107}, 114512 (2023)}.


\bibitem{Flynn:2015mha}
J.~M.~Flynn, T.~Izubuchi, T.~Kawanai, C.~Lehner, A.~Soni, R.~S.~Van de Water, and O.~Witzel,
$B \to \pi \ell \nu$ and $B_s \to K \ell \nu$ form factors and $|V_{ub}|$ from 2+1-flavor lattice QCD with domain-wall light quarks and relativistic heavy quarks,
\href{https://doi.org/10.1103/PhysRevD.91.074510}
{Phys. Rev. D \textbf{91}, 074510 (2015)}.


\bibitem{CLEO:2007vpk}
N.~E.~Adam \textit{et al.} (CLEO Collaboration),
A study of exclusive charmless semileptonic $B$ decay and $|V_{ub}|$,
\href{ https://doi.org/10.1103/PhysRevLett.99.041802}
{Phys. Rev. Lett. \textbf{99}, 041802 (2007)}.


\bibitem{Colquhoun:2022atw}
B.~Colquhoun \textit{et al.} (JLQCD Collaboration),
Form factors of $B\to \pi \ell\nu$ and a determination of $|V_{ub}|$ with M\"obius domain-wall fermions,
\href{https://doi.org/10.1103/PhysRevD.106.054502}
{Phys. Rev. D \textbf{106}, 054502 (2022)}.



\bibitem{FermilabLattice:2015mwy}
J.~A.~Bailey \textit{et al.} (Fermilab Lattice and MILC Collaboration),
$|V_{ub}|$ from $B\to\pi\ell\nu$ decays and (2+1)-flavor lattice QCD,
\href{https://doi.org/10.1103/PhysRevD.92.014024}
{Phys. Rev. D \textbf{92}, 014024 (2015)}.



\bibitem{UTfit:2022hsi}
M.~Bona \textit{et al.} (UTfit Collaboration),
New UTfit analysis of the unitarity triangle in the Cabibbo-Kobayashi-Maskawa scheme,
\href{https://doi.org/10.1007/s12210-023-01137-5}
{Rend. Lincei Sci. Fis. Nat. \textbf{34}, 37 (2023)}.





\end{thebibliography}
\end{document}